\documentclass[a4paper,11pt]{article}
\pdfoutput=1 

\usepackage{jcappub} 
\usepackage{subfig}
\usepackage{graphicx}
\usepackage[utf8]{inputenc}
\usepackage{multirow}
\usepackage{booktabs}
\usepackage{enumitem}

\usepackage[normalem]{ulem}

\usepackage[T1]{fontenc} 
\usepackage{soul}

\title{Hybrid stars with sequential phase transitions: the emergence of the \emph{g}$_2$~mode}

\author[a,b]{M.~C.~Rodr\'iguez,} \author[a,b]{Ignacio~F.~Ranea-Sandoval,} \author[a,b]{M.~Mariani,} \author[a,b]{M.~G.~Orsaria,} \author[a]{G.~Malfatti} \author[c,d]{and O. M. Guilera}

\affiliation[a]{Grupo de Gravitación, Astrofísica y Cosmología, Facultad de Ciencias Astronómicas y Geofísicas, Universidad Nacional de La Plata,Paseo del Bosque S/N, 1900, Argentina}
\affiliation[b]{CONICET, Godoy Cruz 2290, 1425 Buenos Aires, Argentina}
\affiliation[c]{Grupo de Astrofísica Planetaria, Instituto de Astrof\'{\i}sica de La Plata, CCT La Plata-CONICET-UNLP, Paseo del Bosque S/N (1900), La Plata, Argentina.}
\affiliation[d]{Instituto de Astrof\'{\i}sica, Pontificia Universidad Cat\'olica de Chile, Santiago, Chile.}
\emailAdd{mcrodriguez@fcaglp.unlp.edu.ar, iranea@fcaglp.unlp.edu.ar, mmariani@fcaglp.unlp.edu.ar, morsaria@fcaglp.unlp.edu.ar, gmalfatti@fcaglp.unlp.edu.ar, oguilera@fcaglp.unlp.edu.ar}

\abstract{
Neutron stars are the densest objects in the Universe, with $M \sim 1.4 M_{\odot}$ and $R \sim 12$ km, and the equation of state associated to their internal composition is still unknown. The extreme conditions to which matter is subjected inside neutron stars could lead to a phase transition in their inner cores, giving rise to a hybrid compact object. The observation of $2M_{\odot}$ binary pulsars (PSR~J1614-2230, PSR~J0343$+$0432 and PSR~J0740$+$6620) strongly constraints theoretical models of the equation of state. Moreover, the detection of gravitational waves emitted during the binary neutron star merger, GW170817, and its electromagnetic counterpart, GRB170817A, impose additional constraints on the tidal deformability. In this work, we investigate hybrid stars with sequential phase transitions hadron-quark-quark in their cores. We assume that both phase transitions are sharp and analyse the \emph{rapid} and \emph{slow} phase conversion scenarios. For the outer core, we use modern hadronic equations of state. For the inner core we employ the constant speed of sound parametrization for quark matter. We analyze more than 3000 hybrid equations of state, taking into account the recent observational constraints from neutron stars. The effects of hadron-quark-quark phase transitions on the normal oscillation modes $f$ and $g$, are studied under the Cowling relativistic approximation. Our results show that, in the slow conversion regime, a second quark-quark phase transition gives rise to a new $g_2$~mode. We discuss the observational implications of our results associated to the gravitational waves detection and the possibility of detecting hints of sequential phase transitions and the associated $g_2$~mode.
} 

\begin{document}
\maketitle
\flushbottom


\section{Introduction}
\label{sec:intro}

Neutron stars (NSs) are one of the possible compact remnants that can be produced as a result of the gravitational collapse of a massive super-giant star. The observations made over the electromagnetic spectrum suggest that NSs are objects with masses between $1$-$2~M_{\odot}$ and radius between $10$-$14$~km \cite{Ozel2016, Steiner2018}.

Inside NSs, matter is subject to extreme conditions. Central densities are several times higher than nuclear saturation density $n_c\gtrsim\,n_{nuc}=2.3\times10^{14}$~g/cm${}^3$, the typical density of the atomic nucleus \cite{Fridolin1999}. The \emph{equation of state} (EoS) (\emph{i.e} the relation between the pressure, $P$, and energy density, $\epsilon$, of the fluid) is essential to obtain different mass-radius ($M$-$R$) diagrams with relevant features, such as the maximum mass. But the EoS for dense matter inside NSs has yet to be found. For this reason, the  development of different effective models capable of describing the  matter inside NSs is an active area of investigation in theoretical astrophysics. The theoretical search for the EoS along with astronomical observations will help us unravel the mysterious nature and behavior of dense matter. 

Since the properties of matter at nuclear scale are governed by the strong interaction, Quantum Chromodynamics (QCD) is needed for the description of dense matter in NSs. At high densities $(\sim\,2\,n_{nuc})$, QCD predicts a phase transition from the baryonic matter to a state on which the hadrons melt in their constituent particles, the quarks. In this exotic phase, a plasma of free quarks and gluons is formed (see, for example, Ref.~\cite{Orsaria2019} and references therein). In this context, an alternative model to NSs has been proposed: the \emph{hybrid star} (HS) model, compact stars with an inner core of quark matter surrounded by hadronic matter.

Although most of the models of HSs usually consider only one phase transition, from hadronic matter to deconfined quark matter \cite{Ranea2018b, Alford2013, Orsaria2019, Lugones2016, Blaschke2014, Benic2015}, recently an EoS in which two sequential phase transitions occur has been proposed \cite{Alford2017,Li:2019}. The authors of Refs.~\cite{Alford2017,Li:2019} consider a first phase transition from hadronic matter to deconfined quark matter or to the \emph{two-flavor color-superconducting} (2SC) phase, in which quarks up and down pair. Then, a second phase transition took place between this phase and the \emph{color-flavor-locked} (CFL) color superconducting phase in which three flavors of quarks, up, down and strange, form pairs \cite{Alford:2008}. To model such scenario, the authors use a generalization of the \emph{constant speed of sound} (CSS) parametrization for quark matter (see, for example, Refs.~\cite{Alford2013, Ranea2016}). One of their main conclusions is that not only \emph{twin} configurations (\emph{i.e} stable stellar configurations with a given mass but different radius, for more details see section~\ref{subsec:TOV}) can be obtained, but also \emph{triplets}.

The discovery of $\sim 2M_\odot$ pulsars put strong constraints on the EoS above nuclear saturation density and help discard theoretical models unable to reproduce these observations \cite{Demorest2010, antoniadis2013, Cromartie2019}. However, several EoS are still compatible with such constraints \cite{Lattimer2004}. These observational results further highlights the need of accurate measurements not only of masses but radii of these objects. Considering that the determination of the radii of NSs is not simple (see Ref.~\cite{Lattimer2014} and references therein), the search for other alternative quantities that provide us information about the physics of matter inside of these objects becomes essential.

Nowadays, the detection of gravitational waves (GW) emitted by NSs is a turning point in the study of NSs physics, giving rise to the multi-messenger astronomy with GW era \cite{Andersson2011,Lasky:2015,Orsaria2019}. In the last years, the first direct detection of GW coming from the merger of NSs, GW170817, \cite{Abbott2017b} and its electromagnetic counterpart \cite{Abbott2017a} help to restrict the EoS of dense matter putting constraints on the radius of a $1.4M_\odot$ NS \cite{Raithel2018,Annala2018}. Moreover, combining this information with other astronomical observations, recently it has been suggested that quark matter should be present in the inner cores of very massive NSs \cite{quark-NS}. Although a second detection of GW from the merger of NSs, GW190425, was observed, this time no electromagnetic counterpart (from which the lower limit for dimensionless tidal deformability is derived) has yet been detected \cite{Ligo2020}. For this reason, this event does not provide further restrictions to the EoS. Furthermore, new estimates of NSs radii from the NICER experiment, placed at the International Space Station, also help to constraint the EoS. This experiment has estimated the mass and the radius of PSR~J0030+0451 with high precision \cite{Riley2019, Miller2019}.

Apart from the merger of compact objects, another mechanism of GW emission are the non-radial pulsations of NSs (see, for example, Ref.~\cite{Glampedakis:2018} and references therein). In this context, as the source of GW are the oscillations of the matter, the properties and characteristics of the emission modes are intimately related with the EoS that describes the matter inside NSs \cite{Lasky:2015}. Thus, NSs asteroseismology would help us to understand their composition. The main idea of this area of investigation is to parametrize the oscillation frequency, $\omega$, and the damping time, $\tau$, of the oscillation modes as a function of NS macroscopic quantities (for example, mass or radio) in order to obtain an \emph{universal} parametrization, in other words, independent of the EoS. This study field began more than two decades ago with the seminal paper of Andersson and Kokkotas, Ref.~\cite{Andersson1998}. It was first applied to get information from the analysis of the $f$~mode, because it is believed to be the mode most easy to excite in {cataclysmic events such as }the formation of a proto-NS after a supernova event. Today, the importance of the NS asteroseismology lies in determining the dense matter EoS through the estimation of the mass and radius of NSs from GW detections.

The aim of this work is to investigate the effect of sequential phase transitions in the cores of isolated and non-rotating HSs over their non-radial oscillation modes. With this purpose, we will model hybrid EoSs with two sequential phase transitions. For these calculations, we will use an automatized code written in \textsc{Bash}, which includes two \textsc{Fortran} codes, one to calculate the hybrid EoS and the other one to solve the TOV equations of relativistic hydrostatic equilibrium. In this way, we will consider the results presented in Ref.~\cite{extended-stability}, where the authors study the dynamical stability of HSs taking into account the role of phase conversions in the vicinity of a sharp interface, considering two limiting cases: \emph{slow} and \emph{rapid} conversions. This assumption allows us not only to study different families of stable configurations but also to generalize the analysis about the possible existence (or not) of twin or triplet stable branches. Once the stable families in hydrostatic equilibrium are obtained, the $f$ and $g$~oscillation modes will be studied within the relativistic Cowling approximation, using an extended version of the \textsc{Fortran}~CFK code \cite{Ranea2018b}.

From here on, we will use a superscript $i$ ($i=1,2$) to denote the sequential phase transitions, joint conditions and their physical quantities at the respective interfaces.

The paper is organized as follows. In section~\ref{sec:eos}, we present the EoSs that we use to model the hadronic matter in the outer core of HSs. Then, we describe global aspects of quark matter that might be present in the inner core and the simple model used to describe the sequential hadron-quark-quark phase transition. In section~\ref{sec:structure}, we present the relativistic equations of hydrostatic equilibrium together with the one that characterizes the tidal deformability of compact objects. We also describe the non-radial oscillations and the relativistic Cowling approximation used to calculate the oscillation eigenfrequencies. In section~\ref{results}, we show the results obtained not only from solving TOV equation, in which we calculate the mass, radius and tidal deformability, but also from calculating the oscillations modes. Finally, in section~\ref{sec:conclusion}, we present a summary and discussion of this work.


\section{Hybrid equation of state}
\label{sec:eos}

\begin{figure}
    \centering
    \includegraphics[width=0.6\columnwidth]{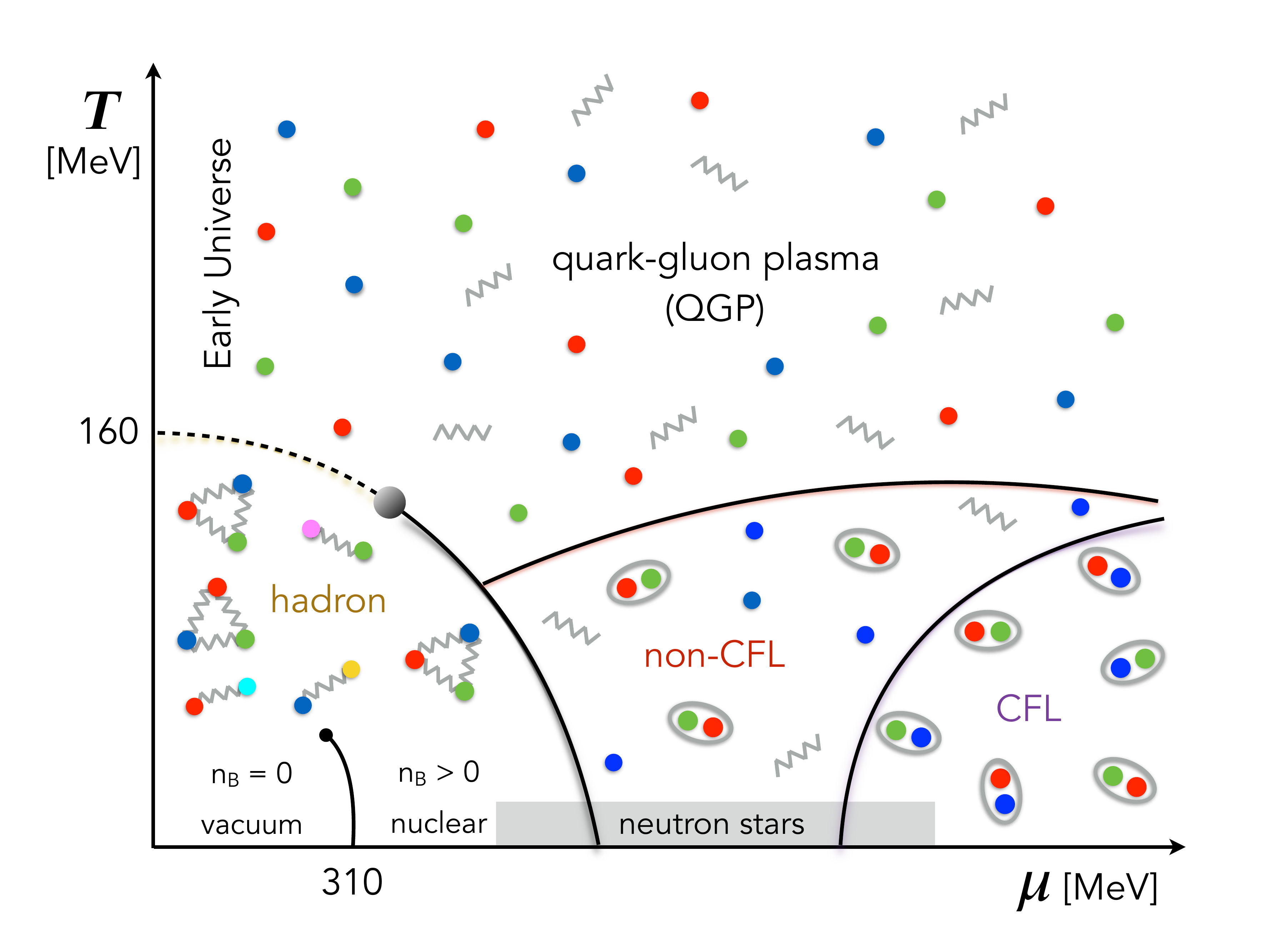}
    \caption{(Color online) Schematic QCD phase diagram. The temperature inside of the HS is considered to be $T=0$. Therefore, the region of HSs lies essentially over the chemical potential axis. As the density increases, matter may change its thermodynamic state going through one or more phase transitions.}
    \label{fig:qcd}
\end{figure}{}

General properties of matter in the inner core of NSs are ruled by strong interactions. Such properties can be analyzed via the so-called QCD phase diagram in the region of low temperatures and intermediate and/or high densities (see figure~\ref{fig:qcd}). The QCD phase diagram is not well established, but both theoretical studies and lattice QCD simulations suggest some common general accepted characteristics \cite{Stephanov:2005, Caines:2017}. For example, at low temperature and high chemical potential, hadrons composed of quarks are thought to dissolve and a hadron-quark first order phase transition occurs. The quark phase could contain free quarks or some non-CFL color superconducting phase, such as 2SC. At even larger values of the chemical potential, another phase transition is expected to occur and quark matter would be in the CFL phase \cite{Alford:2008}. Due to the inherent mathematical difficulties on the QCD treatment, several phenomenological and effective models have been proposed for the description of dense matter: MIT Bag \cite{Chodos1974}, Nambu Jona-Lasino, both local and non-local extension (see, for example, Refs.~\cite{Contrera2010, Orsaria2013, Ranea2019}), Field Correlator Method (see, for example, Refs.~\cite{Nefediev2009, Mariani2017, Mariani2019}). These models reproduce some fundamental properties of QCD and are used for the description of quark matter in the hybrid EoS. 

Figure~\ref{fig:qcd} shows that the area corresponding to NSs may comprise different phases. There are two different more accepted ways in which the hadron-quark phase transition can occur inside HSs. The nature of such phase transition mainly depends on the value of the hadron-quark surface tension, $\sigma_{HQ}$. Due to  the uncertainties related to the value of this quantity and the wide range of theoretical results obtained in the literature, Refs.~\cite{Voskresensky2003, Pinto2012, Lugones2013}, we assume that $\sigma_{HQ}$ is large enough for a sharp phase transition  to be favored. Therefore, we use the \emph{Maxwell construction} to model the hadron-quark-quark phase transitions (see Ref.~\cite{Orsaria2019} and references therein). Besides taking into account results presented in Ref.~\cite{extended-stability}, we consider two limiting scenarios for these sharp phase transitions. These scenarios are characterized through the nucleation time of the phase transition: if the nucleation time is smaller than the oscillation period of the interface under radial perturbations, then we have the slow conversion scenario, otherwise we have the rapid conversion scenario. Since the nucleation times, both for the hadron-quark phase transition and for the quark-quark phase transition considered in our work, are still unknown, we consider both possibilities. As we detail in the next Section, this feature of the phase transitions strongly affects both the dynamical stability of the stellar configurations and the non-radial oscillation modes.

The crust region of HSs is constructed by combining the standard Baym-Pethick Sutherland (BPS) and Baym-Bethe-Pethick (BBP) EoSs \cite{BPS, BBP}. BPS EoS is determined by experimental masses of neutron rich nuclei and it is used for the outer crust, from energy densities $\sim 10^4$ g/cm$^3$ up to the neutron drip energy density $\sim 4 \times 10^{11}$ g/cm$^3$. BBP EoS, a simplified compressible liquid drop model without curvature corrections, is used for the inner crust, extending up to energy densities of $\sim 0.8 \times 10^{14}$ g/cm$^3$. It is worth to mention that despite the crust has almost no effect on the total gravitational mass, nor on the computation of the oscillations modes, it can lead to differences in the NSs radius, affecting the tidal deformability (see, for example, \cite{crust-fortin} and references therein). However, in this work, we will not focus on the crust effects, but on the distinctive characteristics on the frequency spectrum of the oscillation modes that could indicate the presence of an abrupt phase transition in the star.

To describe the hadronic matter for densities smaller or of the order of the nuclear saturation density, we use the Relativistic Mean Field (RMF) EoSs with the parametrizations {DD2}, {GM1L} and {SW4L} \cite{Spinella2017, Typel2010, Spinella:2018dab}. The Lagrangian of these models are given by

\begin{eqnarray}
  \mathcal{L}_{DD2} &=& \sum_{B}\bar{\psi}_B \bigl\{\gamma_\mu [i\partial^\mu - g_{\omega B} \omega^\mu - g_{\rho B} {\boldsymbol{\tau}} \cdot {\boldsymbol{\rho}}^\mu] - [m_B - g_{\sigma B}\sigma] \bigr\} \psi_B \nonumber \\
  && + \frac{1}{2} (\partial_\mu \sigma\partial^\mu \sigma  - m_\sigma^2 \sigma^2)   -  \frac{1}{4}\omega_{\mu\nu} \omega^{\mu\nu} + \frac{1}{2}m_\omega^2\omega_\mu \omega^\mu \nonumber \\
  && + \frac{1}{2}m_\rho^2  {\boldsymbol{\rho\,}}_\mu \cdot {\boldsymbol{\rho\,}}^\mu - \frac{1}{4}  {\boldsymbol{\rho\,}}_{\mu\nu} \cdot {\boldsymbol{\rho\,}}^{\mu\nu} \,,   \\
  \mathcal{L}_{GM1L} &=& \mathcal{L}_{DD2} - \frac{1}{3} \tilde{b}_\sigma m_N (g_{\sigma N} \sigma)^3 - \frac{1}{4} \tilde{c}_\sigma (g_{\sigma N} \sigma)^4\,, \\
 \mathcal{L}_{SW4L} &=& \mathcal{L}_{GM1L} + \sum_{B}\bar{\psi}_B (g_{\sigma^* B}\sigma^*-g_{\phi
      B}\phi^{\mu}) \psi_B + \tfrac{1}{2}\left(\partial_{\mu} \sigma^*\partial^{\mu}\sigma^*-m^2_{\sigma^*}\sigma^{*2}\right) \,\nonumber\\
     && - \tfrac{1}{4}\phi^{\mu\nu}\phi_{\mu\nu}+\tfrac{1}{2}m^2_{\phi}\phi_{\mu}
  \phi^{\mu} \, ,
  \label{eq:Blag}
\end{eqnarray}
where the sum over $B$ implies to sum over the whole baryon octet and the four delta resonances, while the meson fields $\{\sigma,\omega, \boldsymbol{\rho},\sigma^*,\phi\}$ are replaced by their mean field values in the RMF approach. The main difference between {GM1L} and {DD2} is that in the former, only the coupling constants associated with $\boldsymbol{\rho}$ meson, $g_{\rho B}$, are density-dependent, while in the {DD2} case all couplings constants $g_{\omega B}$, $g_{\rho B}$, $g_{\sigma B}$, depend on density, and the $\sigma$ meson self-interactions are not needed. In the {SW4L} case, we consider the inclusion of strange mesons, the scalar $\sigma^*$, and the vectorial $\phi$, and again only the coupling constants associated to the meson $\boldsymbol{\rho}$ depend on density. For {DD2} and {GM1L} parametrizations, the explicit coupling values, as well as the adjusted physical quantities related, can be found in Ref.~\cite{Spinella2017} while the corresponding quantities for the {SW4L} parametrization are those of the Ref.~\cite{Spinella:2018dab}. It is worth clarifying that hadronic EoSs~GM1L and SW4L used in this work satisfy the current observational restrictions for NSs related to mass and radii and have already been used in different previous works  \cite{Ranea2017, Malfatti2019, Mariani2019,Malfatti2020}. 
Despite the fact that not all the HSs constructed with DD2 parametrization  reproduce constraints from GW170817, we will see that a sharp hadron-quark phase transition occurring at low pressure might help to solve this situation (for more details see section~\ref{results}).

To describe quark matter, we use a generalization of the CSS parametrization \cite{Alford2013, Ranea2016}. In this model, the speed of sound inside the HS inner core is normalized by the speed of light and it is assumed constant in each phase, so the EoS is linear. 
\begin{figure}
    \centering
    \includegraphics[width=0.6\columnwidth]{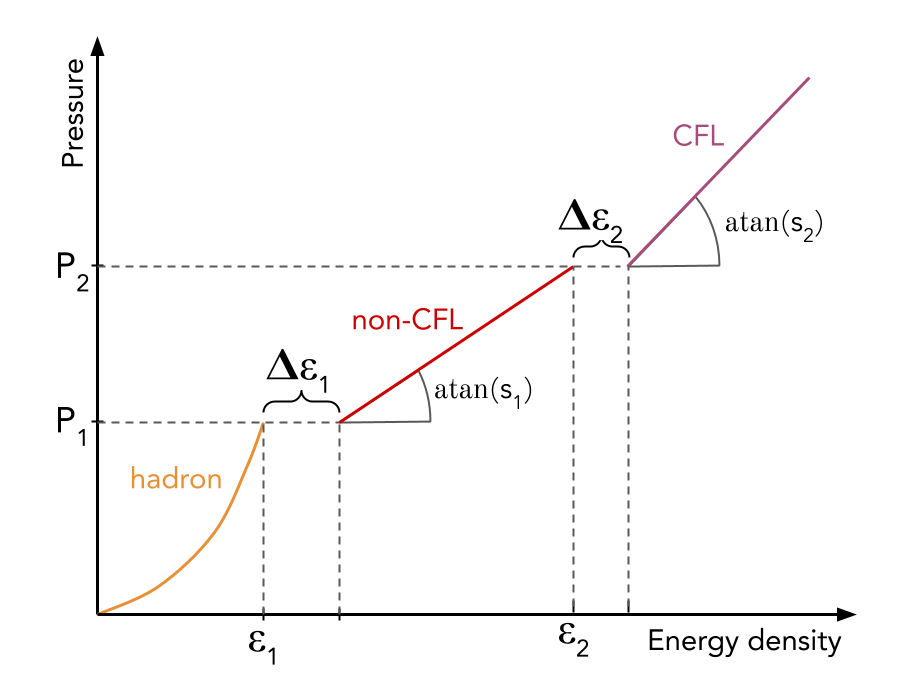}    
    \caption{(Color online) Typical scheme of a hybrid EoS with two abrupt phase transitions. Each color refers to different EoS phases separated by phase transitions, with successive jumps in the energy density $\Delta\epsilon_{1}$ and $\Delta\epsilon_{2}$. The six parameters of the generalized CSS sequential transition model are denoted by $P_1$, $P_2$, $s_1$, $s_2$, $\Delta\epsilon_1$, $\Delta\epsilon_2$. Figure adapted from Ref.~\cite{Alford2017}.}
    \label{fig:eos_alford}
\end{figure}
This generalization describes two sequential sharp phase transitions in terms of six parameters \cite{Alford2017}. 

The quark EoS used in this model has the following behavior:
\begin{equation}\label{eqn:param_css}
    P(\epsilon)= \left\{ \begin{array}{lcc}
             P_{1} & , & \epsilon_{1} < \epsilon < \epsilon_{1}+\Delta\epsilon_{1} \\
             P_{1}+s_{1}[\epsilon-(\epsilon_{1}+\Delta\epsilon_{1})] & , & \epsilon_{1}+\Delta\epsilon_{1} < \epsilon < \epsilon_{2} \\
             P_{2} & , & \epsilon_{2} < \epsilon < \epsilon_{2}+\Delta\epsilon_{2}\\
             P_{2}+s_{2}[\epsilon-(\epsilon_{2}+\Delta\epsilon_{2})] & , & \epsilon < \epsilon_{2}+\Delta\epsilon_{2}\\
             \end{array}
   \right.  
\end{equation}
where $P_{1}$ is the transition pressure between the hadronic and the first phase of quark matter, $P_{2}$ is the transition pressure between the two different states of the quark matter (non-CFL/{2SC} and {CFL}), $\Delta\epsilon_{1}$ is the jump in the energy density for the first phase transition, $\Delta\epsilon_{2}$ is the jump in the energy density for the second phase transition, $s_{1}$ and $s_{2}$ are the square of the speed of sound corresponding to the two different quark phases. For pressures lower than $P_1$, we use one of the three hadronic models previously mentioned. In the figure~\ref{fig:eos_alford} we show a typical hybrid EoS obtained using this parametrization. 

Although in this work we use a piecewise EoS that depends on the combinations of the parameters involved, it is worth to mention the role of the strange quark mass, $m_s$, is crucial for the appearance of a quark-quark phase transition inside a HS if a non-parametric quark EoS is considered. The description of color superconducting phases and the possibility of a phase transition between them, becomes a bit more complex if the quark phase is described by such EoS.

If we assume that very dense matter in the inner core of HSs is made up of weakly interacting quarks, even though such interaction is arbitrarily weak, from an energetic point of view it is more favorable for quarks to form pairs (diquarks) than to be in a quark-gluon plasma. This is due to the Cooper instability, in analogy to Bardeen-Cooper-Schieffer's (BCS) theory of ordinary superconductivity \cite{BCS}. Cooper pairs are bosons and occupy the same lower energy quantum state at zero temperature (Bose condensates). The difference with electrons in ordinary superconductivity is that quarks come in three flavors and three colors in color superconductivity.  As a consequence of diquarks formation, the $SU (3)_c$ color symmetry is broken. As a result of this breakdown, matter inside a HS with diquarks should be not only electrically neutral but should not carry any color. This is achieved by minimizing the quark matter Grand canonical potential with respect the chemical potentials associated with the color charges \cite{Shovkovy_2005}.

The most common patterns of BCS color superconductivity are the 2SC and CFL phases. Figure \ref{fig:qcd} shows the 2SC as the non-CFL phase, in which green and red \emph{up} and \emph{down} quarks pair, while the blue color remains unpaired. All flavors and colors form pairs in the CFL phase. The quarks forming diquarks in 2SC and CFL phases have the same Fermi momentum in magnitude, but with opposite sign. Thus, diquarks have zero net Fermi momentum. This condition is stressed if we consider $m_s \neq 0$ \cite{Alford:2008}. As the density increases in stellar matter, the mass of the strange quark, together with the condition of charge neutrality and $\beta-$equilibrium, produces a mismatch in the Fermi momentum of the pairing quarks. In the 2SC phase only quarks $u$ and $d$ form diquarks (considering $m_u \sim m_d$). If the mismatch in the Fermi momentum increases, the system may experience a phase transition to a more symmetric state, the CFL phase, modeled with $m_u=m_d=m_s=0$ and the same superconducting gap for all pairings in its simplest version.  The quark-quark transition will depend on the competition between $m_s$ and the superconducting gap of each phase, and both are density dependent quantities in more realistic and non-parametric models. A full treatment of how the mass $m_s$ and superconducting gaps vary between different phases could be found in Ref.~\cite{Buballa_2005}.


\section{Hybrid star structure and oscillations} \label{sec:structure}

In this section, we will review some fundamental aspects associated to the theoretical study of static spherical stellar configurations. Moreover, we will discuss some phenomenology related to different families of compact objects and their general properties. We will focus our attention on the concept of \emph{twin} stellar configuration as the hypothetical third branch of stable stars studied in several recent papers \cite{Benic2015, Castillo2017, Montana2019}. Besides, we will present the Cowling approximation that will allow us to study the HSs non-radial oscillations modes. In addition, it is important to mention that throughout this work we use the geometrized unit system in which the gravitational constant, $G$, and the speed of light in vacuum, $c$, are equal to unity.


\subsection{Spherical stellar configurations and stability}
\label{subsec:TOV}

NSs have a typical compactness of about $\beta=M/R \sim 0.15$, thus it is required to work in the framework of General Relativity to describe properly its structure. Besides, most NSs are, with high level of precision, spherical objects \cite{Bhattacharyya:2020paf,Abbott:2020lqk}. For these reasons, it is valid to work in a spherically symmetric space-time characterized by the following line element,
\begin{equation*}
    {\rm d}s^2= -\exp{(\nu (r))} {\rm d}t^2 + \left( 1- \frac{2m(r)}{r}\right)^{-1} {\rm d} r^2 + r^2({\rm d}\theta ^2 + \sin ^2 \theta {\rm d}\phi ^2) \, ,
\end{equation*}
where functions $\nu (r)$ and $m(r)$ are obtained after solving Einstein field equations. Starting from this line element, the equations to obtain the structure of these compact objects were obtained by Tolman \cite{tolman} and Oppenheimer and Volkoff \cite{OV}. They are commonly referred to as the TOV equations, and are given by:
\begin{eqnarray} \label{toveq}
\frac{{\rm d}P (r)}{{\rm d}r}&=&- \frac{\epsilon (r) m(r)}{r^2}\left[1 + \frac{P(r)}{\epsilon (r)} \right] \left[ 1+\frac{4\pi r^3 P(r)}{m(r)}\right] \left[1 - \frac{2m(r)}{r} \right]^{-1} \, , \nonumber \\
\frac{{\rm d}m(r)}{{\rm d}r}&=&4\pi r^2 \epsilon (r) \, , \\
\frac{{\rm d} \nu (r)}{{\rm d}r} &=& -\left( \frac{2}{P(r) + \epsilon (r)}\right)\frac{{\rm d} P(r)}{{\rm d}r} \, , \nonumber
\end{eqnarray}
where $m(r)$ is the total mass enclosed at a radius $r$ and $P$ and $\epsilon$ are the pressure and the energy density, respectively. To be able to solve this system of differential equations we need not only to establish an EoS and to select a central energy density, $\epsilon _c$, but also to set the proper boundary conditions:
\begin{eqnarray*}
m(0) &=& 0 \, ,\\
P(R) &=& 0 \, , \\
{\rm exp}\left(\nu (R)\right) &=& 1 - \frac{2M}{R} \, ,
\end{eqnarray*}
where $M$ is the total mass of the star and $R$ its radius.

After solving equations~\eqref{toveq}, for a given set of central energy densities, we obtain the so-called mass-radius ($M$-$R$) and mass-central energy density ($M$-$\epsilon_c$) relationships for a given EoS. In addition, both the pressure and energy density profiles can be obtained for each star composing the family of stellar configurations. $M$-$R$ and $M$-$\epsilon_c$ curves contain crucial information that can be compared with astronomical observations of NSs mass and radius. Through this process, we can validate or discard theoretical models of NSs' EoSs \cite{Orsaria2019}.

Besides the mass and radius of the compact object, we can calculate the dimensionless tidal deformability, $\Lambda$. This quantity contains information related to the (in)ability of an external gravitational field to induce a mass-quadrupole momentum on a given compact object. Restrictions on this quantity were provided based on the GW data from the pre-merger stage of the BNS mergers. For quadrupolar perturbations, $\Lambda$ is related to the Love number, $k_2$, that can be obtained by solving, together with the TOV equations~\eqref{toveq}, an additional differential equation:
\begin{eqnarray} \label{eqz}
r\frac{{\rm d}\zeta (r)}{{\rm d}r} &+& \zeta (r)^2 + \zeta (r)\frac{1+4\pi r^2 \left[P(r) + \epsilon (r)\right]}{1-\frac{2m(r)}{r}} \nonumber \\
&-& \frac{6 - 4\pi r^2 {\rm exp}\left(\lambda (r)\right)\left[5\epsilon (r) + 9P(r) + \frac{\epsilon (r) + P(r)}{{\rm d}P/{\rm d}\epsilon}\right]}{1-\frac{2m(r)}{r}} -\left(r\frac{{\rm d}\nu (r)}{{\rm d}r}\right)^2 = 0 \, , 
\end{eqnarray}
with boundary condition $\zeta (0) = 2$. Moreover, for EoSs with sharp discontinuities at radii $r=r^i_{\rm t}$, the additional junction condition
\begin{equation}
\zeta ({r^i _{\rm t}}^+) - \zeta ({r^i_{\rm t}}^-) = \frac{4\pi {r^i_{\rm t}}^3 \left[\epsilon ({r^i_{\rm t}}^+) - \epsilon ({r^i_{\rm t}}^-)\right]}{m({r^i_{\rm t}}) + 4\pi (r^i_{\rm t})^3P(r^i_{\rm t})} \, ,
\end{equation}
must be imposed \cite{2020PhRvD.102b8501T}. After solving equation~\eqref{eqz}, we calculate the $k_2$ Love number using the following expression:
\begin{eqnarray}
k_2 &=& {\frac{8}{5}\beta^5(1-2\beta)^2[2+2\beta(\zeta-1)-\zeta]}\times \Big[2\beta[6-3\zeta+3\beta(5\zeta-8)] \\
&+& 4\beta^3[13-11\zeta+\beta(3\zeta-2)+2\beta^2(\zeta+1)] +  3(1-2\beta)^2[2-\zeta+2\beta(\zeta-1)]\ln(1-2\beta)\Big]^{-1} \, , \nonumber
\end{eqnarray}
where $\zeta \equiv \zeta(R)$. Then, the dimensionless tidal deformability is obtained using
\begin{equation} \label{dimtidal}
    \Lambda = \frac{2}{3}k_2\left(\frac{R}{M}\right)^5 \, .
\end{equation}

It is worth to mention that the individual values of the dimensionless tidal deformabilities for the BNS system of GW170817 have been constrained \cite{Raithel2018,Annala2018}. These observations became extremely useful to discard theoretical models of NSs' EoSs. 

Regarding the dynamical stability of the stellar configurations, in Ref.~\cite{extended-stability} the authors study the response of HSs to radial perturbations and find that, within the slow conversion scenario, extended branches of stable stellar configurations might appear, even when the total mass of each star is not an increasing function of the central energy density. In contrast, for \emph{rapid} conversions, the stability criterion is valid only for the $\partial M  / \partial \epsilon _c > 0$ branches. The details of this calculation, that implies considering the joint condition at the phase transition interface, can be found in Ref.~\cite{extended-stability}.

At this point we have to mention the possibility of the existence of \emph{standard} twin stars and how it would change the traditional $M$-$R$ relationship. By \emph{standard} we mean a third stable family of compact objects under the usual stability criterion, given by the assumption of a rapid conversion. The theoretical possibility of this third stable family of compact objects was proven by Gerlach in the late 1960s \cite{Gerlach:1968}. He found that the necessary condition for this family to exist is a large discontinuity in the speed of sound. In this scenario, after a branch of standard unstable stellar configurations, a new branch of stable stars appears. Such theoretical feature is also present in several works related to NSs with a hadron-quark phase transition in their cores \cite{Schertler:2000,Banik:2002,Ranea2017,Sieniawska:2018,Alvarez-Castillo:2018}. In contrast, under the slow conversion case, the mentioned unstable branch could become stable and then, the usual main and twin branches get connected by stable configurations giving rise to triplets. The occurrence of standard twin configurations is a clear indicator of the existence of a hadron-quark phase transition in the core of compact objects. However, it would not necessarily be useful to determine the nature of the phase transition since twins are present both in models where a sharp hadron-quark phase transition occurs and in those where a mixed phase appears \cite{pasta-twins1,pasta-twins2}. Astrophysical relevance of twin stellar configurations has gained attention since data from GW170817 (and its electromagnetic counterpart) and from the NICER observation of PSR~J0030+0451 began to be in tension (see panel~\hyperref[fig:mraio-mrho]{3.(a)} for more details). Clearly, the detection of GW emitted by isolated compact objects might be a key point to shed some light into understanding the behavior of matter inside these extreme objects.


\subsection{Non-radial oscillations}
\label{subsec:nonradial}

One possible source of GW are isolated NSs, which can be perturbed by many factors (oscillations of a binary NS post-merger remnant, glitch/anti-glitch activity, mountains, magneto-elastic oscillations in magnetars, etc. \cite{Glampedakis:2018}) and, to restore balance, they began to oscillate. Non-radial oscillation of non rotating stars emits GW as long as they have non zero quadrupolar moments, \emph{i.e}, the perturbation associated with $l \geq 2$ is not null. The energy is expected to be released, mostly, through excitation of a few of these quasi-normal modes and in general, the strongest emission occurs for quadrupolar modes, corresponding to $l=2$ \cite{kk-and-2002}. Hence, we will focus our attention on this case. 

The non-radial oscillation modes, not only \emph{fluid} modes but also space-time modes, are known as \emph{Quasi-normal modes} (QNMs) because they are energy-dissipation modes. The eigenfrequency of these modes are complex, whose real and imaginary part are related to the oscillation frequency and damping time, respectively. Therefore, such modes describe perturbation fields decreasing in time.

The fluid modes ($f$, $p$, $g$, etc.)  are related to perturbations in the fluid that excite spacetime oscillations \cite{Kokkotas1999}. On the other hand, a different family of modes, like the $w$, or space-time modes ($w$, $wII$, etc.) are oscillations of the metric that hardly excite the fluid motion \cite{Kokkotas1992}. The most important fluid modes related to GW emission are the fundamental modes ($f$), the pressure modes ($p$), and the gravity modes ($g$). The $p$~modes have higher frequencies than the $g$~modes, and their families are separated by the $f$~modes. Moreover, there is only one $f$~mode for each index $l$ in the spherical harmonic, while there are infinite $p$ and $g$~modes \cite{Kokkotas1999}.

The presence of $g$~modes in cold and non rotating stars, indicates that exist an abrupt phase transition in the interior of the star, because these modes are completely inhibited if the EoS is continuous \cite{Finn1987}. Also, it is important to stress that the $g$~modes associated with discontinuities in the EoS only exist in the slow conversion scenario: $g$~modes does not appear if the conversion is rapid \cite{tonetto2020}.

On the other hand, there are other effects that excite different families of $g$~modes that may differ from the ones produced by discontinuities in the EoS. In Refs.~\cite{Miniutti2003,Lugones2016}; for example, the authors analyzed the $g$~modes generated by considering effects as superfluidity, rotation, and also the existence of intense magnetic fields in the interior of NSs. Once presented our results, we discuss this with more details in section~\ref{sec:conclusion}.


\subsubsection{Relativistic Cowling approximation}
\label{ssubsec:cowling}

In order to study the non radial oscillation modes of HSs, we use the relativistic Cowling approximation \cite{Finn1988}. Within this approximation, the perturbations of the background space-time are neglected and only the fluid perturbations are considered. Nevertheless, the oscillation frequencies obtained using this method differ only by $20\%$ from those obtained using the General Relativity linearized equations (see Refs.~\cite{VasquezFlores2014, Chirenti2015, Ranea2018b}, and references therein). The error related to the fundamental mode is reduced as the compactness of the NS increase, but it is never larger than $20\%$ \cite{Chirenti2015}. On the other hand, if we focus on the error associated to the $g$~modes, the difference is less than $10\%$ \cite{Sotani2002,tonetto2020}. Therefore, for the $g$~modes, the results obtained using the relativistic Cowling approximation are qualitatively and quantitatively acceptable.

When the metric perturbations are negligible, the set of differential equation necessary to study the oscillation modes is simplified  considerably. In the relativistic Cowling approximation only the fluid modes can be studied. Also, since there is no energy dissipation by GW emission, the damping times of the QNMs can not be studied. Nevertheless, it is not a shortcoming since we aim to focus on the normal stellar oscillations modes which can be studied through this approximation without problems. Within this theoretical framework, the equations needed to obtain the mode oscillation frequencies are:
\begin{equation}\label{eqn:dwdr}
    \frac{{\rm d}W(r)}{{\rm d}r} = \frac{{\rm d}\epsilon}{{\rm d}P} \left[\omega^2\frac{r^{2}}{\sqrt{1-\frac{2m(r)}{r}}}e^{-\nu(r)} V(r) + \frac{1}{2}\frac{{\rm d}\nu(r)}{{\rm d}r} W(r)\right] - \frac{l(l + 1)}{\sqrt{1-\frac{2m(r)}{r}}} V (r)  \ ,
\end{equation}{}
\begin{equation}\label{eqn:dvdr}
    \frac{{\rm d}V (r)}{{\rm d}r} = \frac{{\rm d}\nu(r)}{{\rm d}r} V (r) - \frac{1}{\sqrt{r-2m(r)}}W(r)  \ ,
\end{equation}
where $\omega$ is the oscillation frequency and $W(r)$ and $V(r)$ are functions of $r$ that characterize the fluid perturbation \cite{Detweiler1985,Ranea2018b}.

To solve this set of coupled differential equations, \eqref{eqn:dwdr} and \eqref{eqn:dvdr}, we need to impose boundary conditions in two points inside the star. At the center ($r \sim 0$), the behavior of the solutions are of the form
\begin{equation}\label{eqn:borde1}
    W(r) \sim r^{l+1}  \ , \  V (r) \sim -l^{-1}r^{l} \ ,
\end{equation}{}
and at the stellar surface, the Lagrangian perturbation to the pressure has to be zero,
\begin{equation}\label{eqn:borde2}
    \Delta P(R) = 0 \ .
\end{equation}{}
After some algebra, condition \eqref{eqn:borde2} can be written as
\begin{equation}\label{eqn:borde}
    \omega^{2}\left[{\sqrt{1-\frac{2M}{R}}}\right]^{-\frac{3}{2}}V(R) + \frac{1}{R^{2}} W(R)=0  \ .
\end{equation}
The values of $\omega ^2$ that satisfy the equation~\eqref{eqn:borde} are the eigenfrequencies we seek to obtain \cite{Sotani2002,Sotani2011}.

This method can be generalized to the case in which the EoS has one or more discontinuities at $r = r^i_t$. In this case, additional junction conditions at the surface of each discontinuity have to be considered. These continuity conditions for $W$ and $\Delta P$ can be written in terms of the variables $W$ and $V$ in the following way:
\begin{equation}\label{eqn:pegado3}
    W^i_{+}=W^i_{-} \ ,
\end{equation}
\begin{equation}\label{eqn:pegado4}
    V^i_{+} = \frac{e^{\nu}}{\omega ^2 {r^i_t}^{2}} \left(1 - \frac{2m}{r^i_t}\right) \times \left( \frac{\epsilon^i_{-}+P}{\epsilon^i_{+}+P}[\omega^{2}{r^i_{t}}^{2}e^{-\nu}\left(1-\frac{2m}{r^i_t}\right)^{-1}V^i_{-} + \nu' W^i_{-}]-\nu' W^i_{+}\right) \ ,
\end{equation}
where the subindex $-$ ($+$) corresponds to quantities before (after) the phase transition. Note that, in absence of discontinuities, the function $V(r)$, becomes continuous and  
$V_+ = V_-$.

These calculations are performed using an extended version of the CFK code \cite{Ranea2018b}. This new version is able to obtain oscillation modes of stellar configurations with two sharp phase transitions. Thus, given an initial frequency, $\omega_0^2$, the system of equations~\eqref{eqn:dwdr} and \eqref{eqn:dvdr} is solved using a Runge-Kutta-Fehlberg integrator. In the case of HSs with two phase transitions, we first integrate from the center of the star up to the first transition. Then, we integrate again up to the second transition, using the continuity conditions given by equations~\eqref{eqn:pegado3} and \eqref{eqn:pegado4} at the first transition, to obtain the corresponding initial conditions for this region. Finally, we integrate up to the surface of star, using equations~\eqref{eqn:pegado3} and \eqref{eqn:pegado4} at the second transition to compute the initial conditions of this third region of integration. Due to the stiff nature of the eigenfrequency equation root-finding problem, we use a Newton-Raphson algorithm coupled with Ridders’ method to obtain a corrected frequency, and we repeat the aforementioned procedure until equation~\eqref{eqn:borde} is fulfilled.


\section{Results}
\label{results}
To analyze the structure and stability of the NS models and the frequencies of their oscillation modes, we construct a large amount of different hybrid EoSs. As it has been already mentioned in section~\ref{sec:eos}, for the crust we use the BPS-BBP model and for the hadronic phase we used the DD2, GM1L and SW4L parametrizations of the RMF, with density dependent coupling constants. For the sequential quark phases, the parameters have been chosen taking into account two cases: \emph{stiff} and \emph{soft}. For the \emph{stiff} case, we consider an extended range of the parameter values used in Ref.~\cite{Alford2017}, $s_1=0.7$ and $s_2=1.0$. For the \emph{soft} case, we take lower values of the square speed of sound, $s_1=0.3$ and $s_2=0.5$, consistent with non-parametric quark EoSs \cite{Ranea2017, Ranea2019, Mariani2019}. For the rest of the parameters, $P_1$, $P_2$, $\Delta \epsilon_1$ and $\Delta \epsilon_2$, we consider the following range of values:
\begin{itemize}
    \item[] $38~\textrm{MeV/fm}^{3} \leq P_1 \leq  410~\textrm{MeV/fm}^{3} \, , \ 200~\textrm{MeV/fm}^{3} \leq \Delta \epsilon_1 \leq 1000~\textrm{MeV/fm}^{3}$\, ,
    \item[] $125~\textrm{MeV/fm}^{3} \leq P_2 \leq  610~\textrm{MeV/fm}^{3} \, , \ 22~\textrm{MeV/fm}^{3} \leq \Delta \epsilon_2 \leq 38~\textrm{MeV/fm}^{3}$\, ,
\end{itemize}
generating about $3400$ hybrid EoSs by combining the parameters in different ways. Then, after solving the TOV equations for each EoS, we obtain the mass, $M$, and radius, $R$, of the stellar models. Analyzing these results, we select a set of eight different representative EoSs whose $M$-$R$ curves have a star of maximum mass larger than $2.05M_{\odot}$, with twins or triplet configurations, if exists, and with connected and disconnected stability branches. These selected models are presented, described and labeled in table~\ref{tabla:param_selec}.

\begin{table}[t!]
\centering
\begin{tabular}{c c c c c c c c}
\toprule
\multirow{2}{*}{\begin{tabular}[c]{@{}l@{}} \\ Hyb. EoS \end{tabular}} & \multirow{2}{*}{\begin{tabular}[c]{@{}l@{}} \\ Had. EoS \end{tabular}} & \multicolumn{6}{c}{Quark EoS} \vspace{0.1cm}  \\
& & $P_1[{\rm \frac{MeV}{fm^{3}}}]$ & $P_2[{\rm \frac{MeV}{fm^{3}}}]$ & $\Delta \epsilon _1[{\rm \frac{MeV}{fm^{3}}}]$ & $\Delta \epsilon _2[{\rm \frac{MeV}{fm^{3}}}]$ & $s_1$ & $s_2$ \vspace{0.2cm} \\
\hline
\midrule
1 & DD2 & 38 & 150 & 220 & 22 & 0.7 & 1.0 \\ \midrule
2 & DD2 & 50 & 150 & 220 & 38 & 0.7 & 1.0 \\ \midrule
3 & DD2 & 125 & 150 & 220 & 38 & 0.7 & 1.0 \\ \midrule
4 & DD2 & 250 & 400 & 200 & 30 & 0.3 & 0.3 \\ \midrule
5 & GM1L & 38 & 150 & 220 & 22 & 0.7 & 1.0 \\ \midrule
6 & SW4L & 350 & 380 & 220 & 30 & 0.7 & 1.0 \\ \midrule
7 & SW4L & 350 & 450 & 220 & 22 & 0.7 & 1.0 \\ \midrule
8 & SW4L & 410 & 570 & 220 & 38 & 0.7 & 1.0 \\ 
\bottomrule
\end{tabular}
\caption{Details of the eight selected hybrid EoSs: number label, hadronic EoS and the six parameters of the sequential CSS model for the quark phase.}
\label{tabla:param_selec}
\end{table}

\begin{figure}[t!]
    \centering
    \includegraphics[width=0.49\columnwidth]{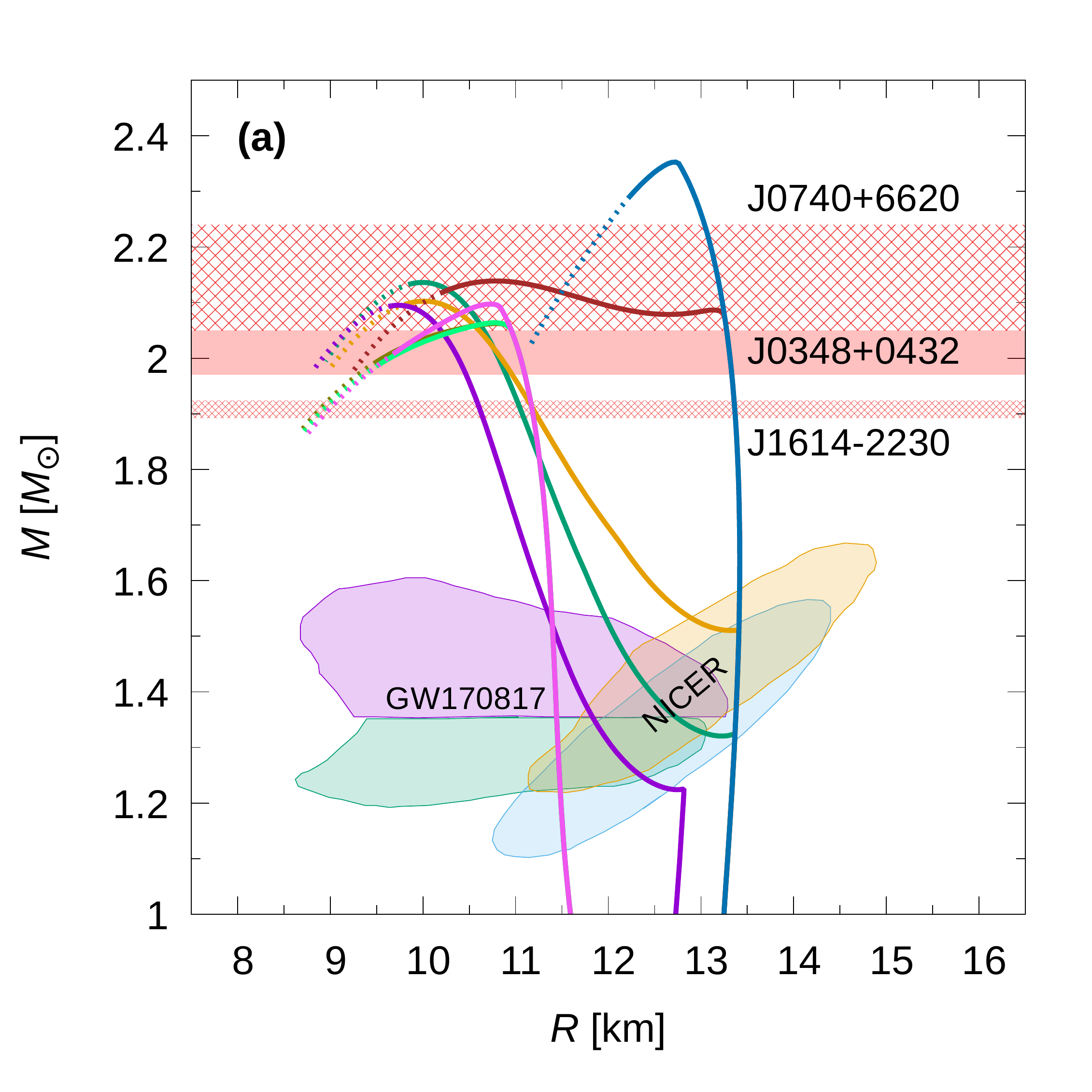}
    \includegraphics[width=0.49\columnwidth]{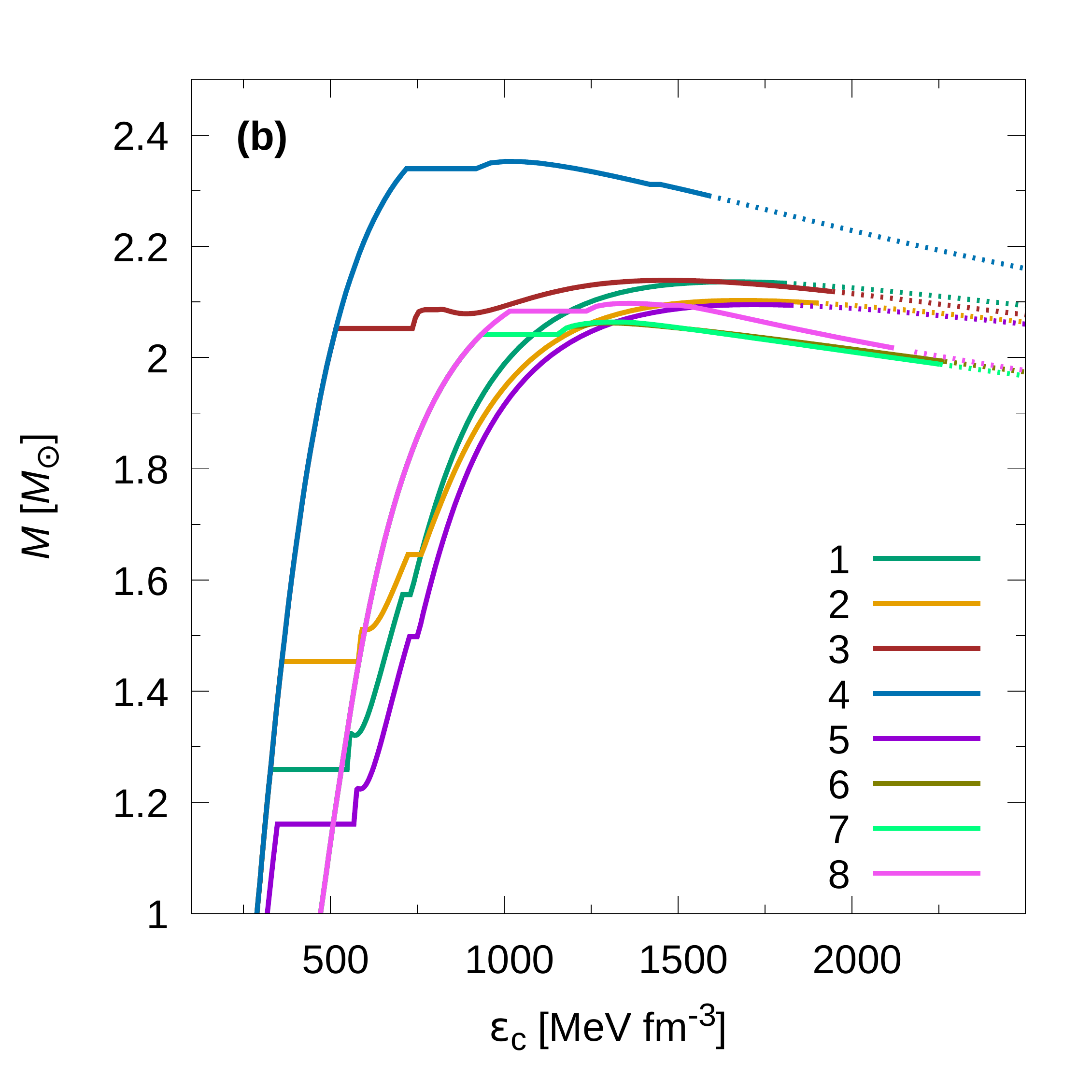}
    \caption{(Color online) $M$-$R$ relationship is presented for the selected EoSs, together with five observational constraints, panel (a). The colored regions correspond to the constraints imposed by the GW170817 event and NICER observations. The horizontal bars correspond to the constraints imposed by J0740+6620, J0348+0432 y J1614-2230. On the right, panel (b), the $M(\epsilon_c)$ relationship is presented for the same EoSs. In both figures, the solid lines correspond to stable branches (if we consider slow conversion between phases) while the dotted lines correspond to unstable branches. For rapid phase transitions stable configurations are only those for which $\partial M/\partial \epsilon _c >0$. Each color represent different hybrid EoS, whose references are in table~\ref{tabla:param_selec}.}
    \label{fig:mraio-mrho}
\end{figure}

In figure~\ref{fig:mraio-mrho} we show our TOV integration results for the representative hybrid EoSs of table~\ref{tabla:param_selec}, where each color represents each EoS. Under the slow conversion scenario, in both panels, the solid lines correspond to stable branches, while the dotted lines correspond to unstable branches. For the \emph{rapid} conversion case, the stable branches correspond, as usual, to the regions where $\partial M / \partial \epsilon_c > 0$ and $\partial M / \partial R < 0$. In panel~\hyperref[fig:mraio-mrho]{3.(a)}, we show our results for the relation $M$-$R$ and the constraints from current observations. In panel~\hyperref[fig:mraio-mrho]{3.(b)}, we also show the relation $M(\epsilon_c)$, where it can be seen how the jumps in $\epsilon_c$ in each curve indicate the abrupt sequential phase transitions. The stability vanishes for $\partial M / \partial \epsilon_c < 0$ when rapid conversions are considered, but it is extended until the \emph{terminal mass}- the last stable configuration of the extended branch- if the slow conversion between phases scheme is assumed.

As it can be seen from panel~\hyperref[fig:mraio-mrho]{3.(a)}, all the selected EoSs satisfy the constraints from J0740+6620, J0348+0432 and J1614-2230. The restrictions imposed by GW170817 and NICER for masses and radii are not satisfied by all the selected hybrid models. While all the GM1L and SW4L hybrid EoSs satisfy these restrictions, only one DD2 hybrid EoS does it. For this EoS, EoS~1, there is an early phase transition between the DD2 parametrization and quark matter. The purely hadronic DD2~NSs and  the DD2~HSs with a late phase transition do not fulfill GW170817 constraints.  

When we work considering that both phase transitions are rapid, a few standard twin configurations appear for EoS~1, 3 and 5. In general, even having explored the parameter space in a systematic and exhaustive way, our results show that within this CSS sequential model, the existence of twins satisfying the latest observational restrictions of NSs is marginal and the existence of triple stars, null. It also should be noted that, for both EoS~1 and 5, the star configurations satisfying the observational constraints are in a disconnected hybrid stable branch. This situation changes when slow phase transitions are considered and extended stable branches appear. Within this theoretical scenario, high-mass \emph{slow twins} (we use this name to distinguish them from the standard twins) are a common feature (see, EoS 3, 4, 7 and 8 in figure \ref{fig:mraio-mrho}). Moreover, it is important to note that \emph{slow triplets} also appear. This situation is particularly visible in EoS 3. One interesting case is the one of EoS 4, in which no stable stellar configurations with two phase transitions are present if we consider the rapid conversion scenario. This situation changes when phase transitions are assumed to be slow, where an extended stable branch that reach the second phase transition appears.

\begin{figure}[t!]
    \centering
    \includegraphics[width=0.6\columnwidth]{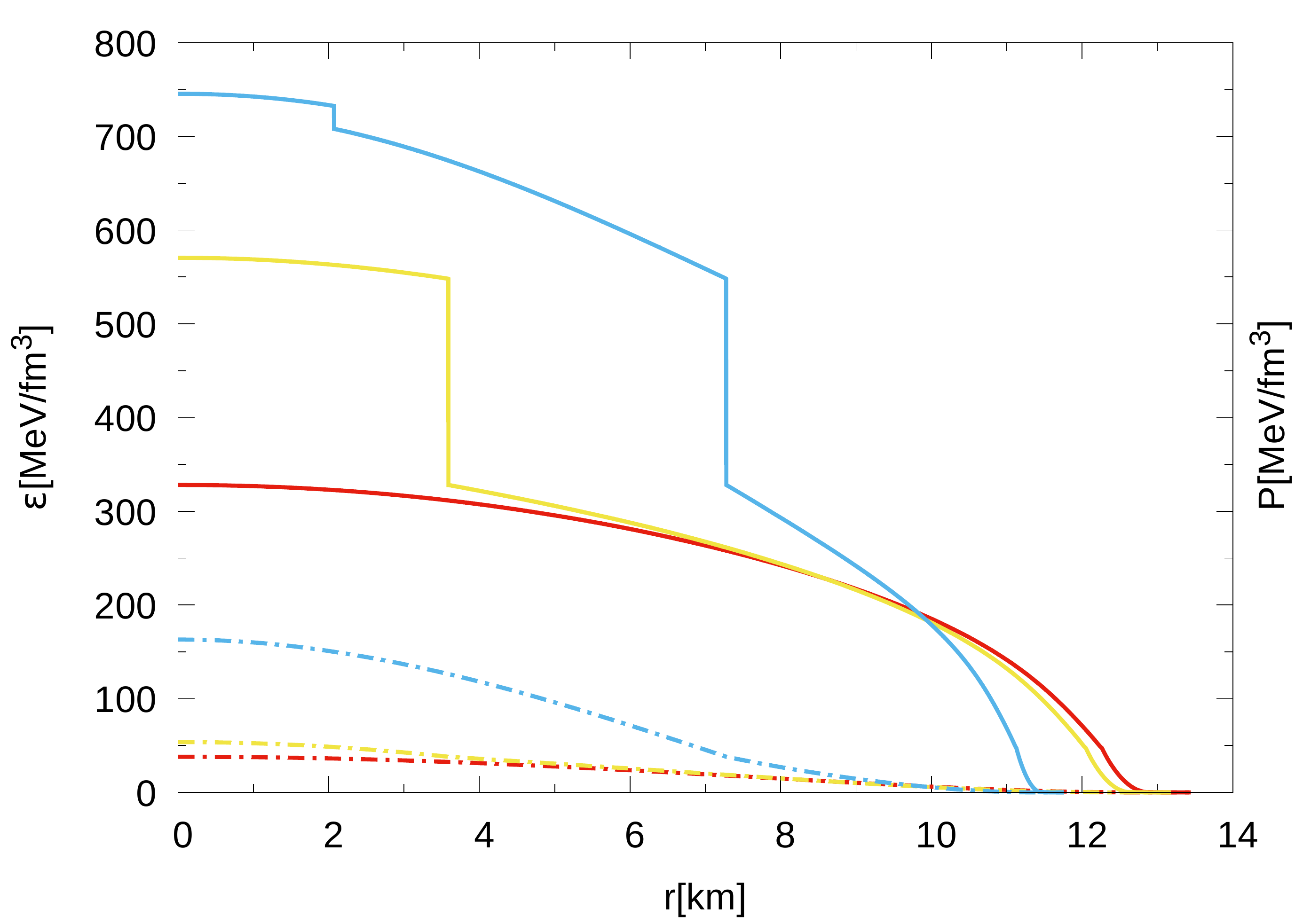}
    \caption{(Color online) In continuous lines, we present the energy density profile, $\epsilon(r)$, for stars with hadronic EoS (red), one phase transition (yellow) and two phase transitions (light blue). With the same colors, but using dashed and dotted lines, we present the pressure profiles, $P(r)$, continuous in all cases. Both quantities, energy density and pressure, are in the same scale. In this schematic figure, it can be seen the radius at which the phase transition occurs for HSs with one and two phase transitions.}
    \label{fig:perfiles}
\end{figure}

Additionally, in figure~\ref{fig:perfiles}, we show the energy density (continuous lines) and pressure (dashed-dotted lines) internal profiles, as function of the stellar radius, for the three types of stellar configurations that could result from considering the sequential hybrid EoS presented in section~\ref{sec:eos}. In this schematic figure, the specific details of each EoSs are not relevant since we aim only to show the qualitative features of each EoS profile. We show, in red color, a purely hadronic star, in yellow, a hybrid star with one phase transition and, in light blue, a hybrid star with two phase transitions. For the hadronic star, both the pressure and the energy density are continuous functions of the radius. The hybrid stellar configurations with one (two) phase transition(s) show one (two) discontinuity(ies) in the energy density. In this sense, the stellar models we present have, comparing these cases, similar radius but radically different internal compositions. Also, in purely hadronic stars, not only the pressure but also the energy density are lower than in HSs. As the compactness of the star increases, the pressure and the energy density inside the star increase too.

\begin{figure}[t!]
    \centering
    \includegraphics[width=0.7\columnwidth]{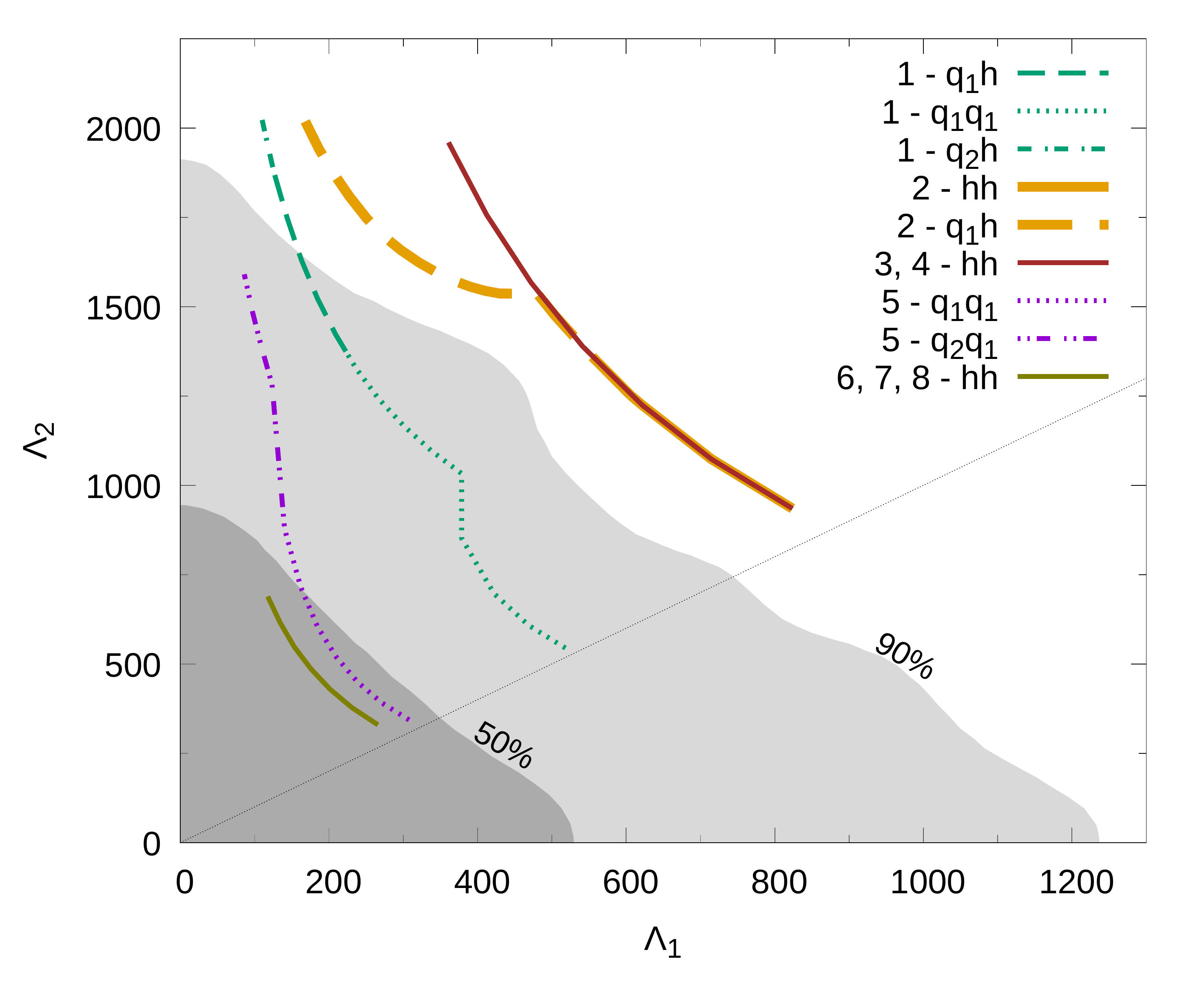}
    \caption{(Color online) Dimensionless tidal deformabilities in the $\Lambda_1-\Lambda_1$ plane for the EoSs of table~\ref{tabla:param_selec}. Each axis corresponds to one of the two stars of the NS-binary system, considering the constraint imposed by GW170817. The gray regions correspond to the $90\%$ and $50\%$ probability contour given by this event. As the keys of the figure shows, the overlapped EoSs are presented with only one color. The dash style of each curve indicates the number of phase transitions that are present inside each of the two NS of the system: \emph{$h$} stands for pure hadronic NS, \emph{$q_1$} for HS with only one phase transition and \emph{$q_2$} for HS with two phase transitions.}
    \label{fig:tidales2}
\end{figure}

In figure \ref{fig:tidales2}, we show the individual dimensionless tidal deformabilities for compact objects with masses in the range consistent with the (astrophysically more plausible) low-spin case for GW170817 \cite{Abbott2017a}. For this figure, we do not differentiate between the rapid and slow scenarios in order to keep the figure simple an clear, since there are no qualitative differences in the implications of both results. The sequential phase transition model that we are working with opens several possible scenarios in which not only NSs might have merged but also situations in which a NS and a HS merged and cases in which the merging objects were two HSs. The cases in which NSs merge are labeled with hh, those in which a HS merge are indicated with ${\rm q}_i$ where $i=1,2$ indicates how many phase transitions occur in the HS interior. Finally, the cases in which two HSs merge are labeled with $\rm{q}_i\rm{q}_j$. We also present the 50$\%$ and 90$\%$ confidence level curves of Ref.~\cite{Abbott2017a}. The general effect of the appearance of quark matter is to produce lower values of the dimensionless tidal deformability, these results are in agreement with those obtained in Ref.~\cite{Li:2019}. We see that the models with DD2 hybrid EoSs are only able to reproduce these observations if one of the objects that merge in GW170817 is a HS and the transition pressure from hadronic matter to non-CFL quark matter is low as in EoS~1 (see, table~\ref{tabla:param_selec} for details). If one consider the other two hadronic parametrizations, GM1L and SW4L, they are consistent with the available astronomical constraints both for purely hadronic stars as for hybrid ones. Parametrization SW4L produces more compact hadronic stars, for this reason they are in better agreement with GW data.

\begin{figure}[htp]
  \centering
  \includegraphics[width=1.0\columnwidth]{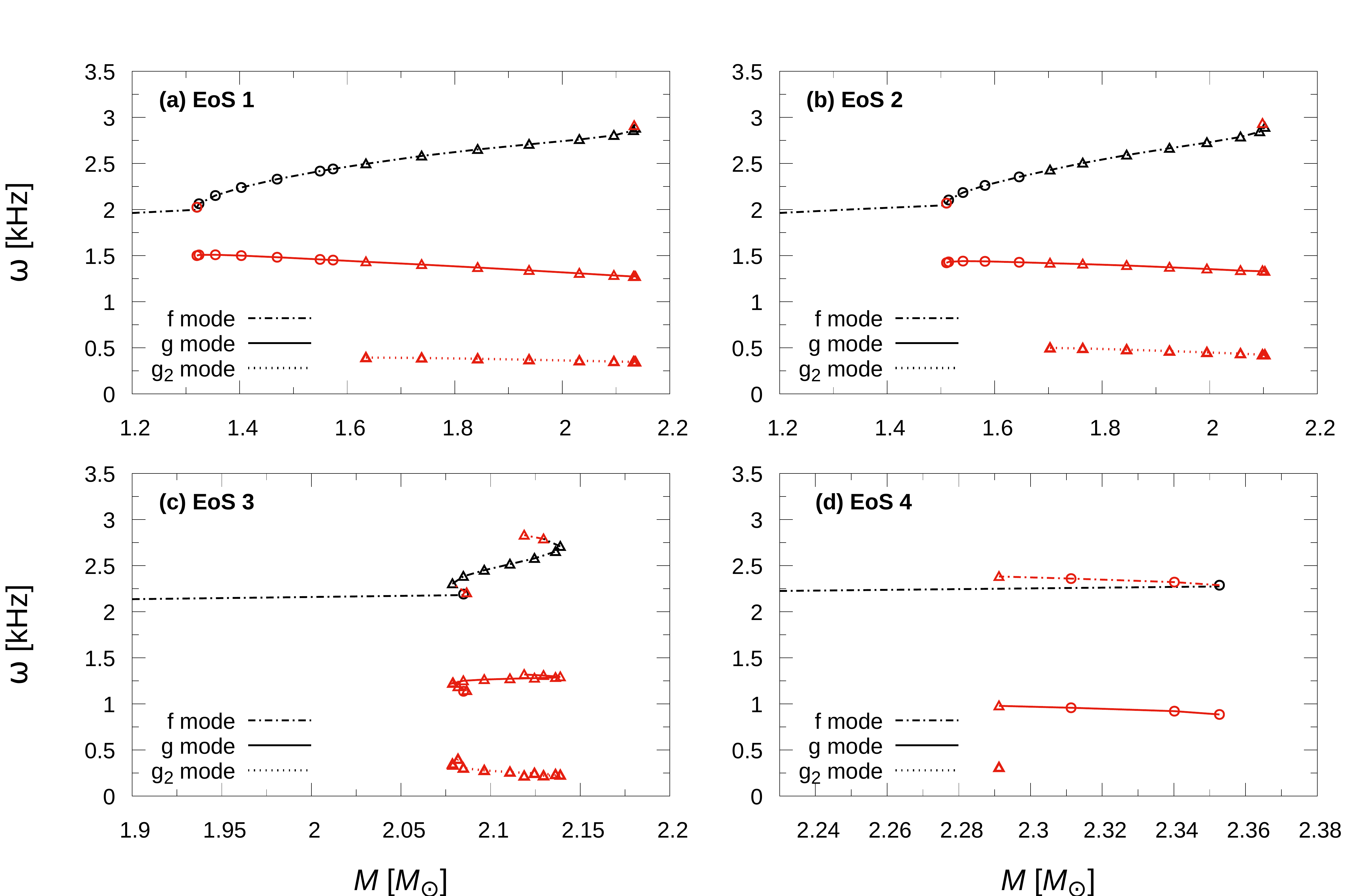}
  \caption{(Color online) Frequencies of the $f$, $g$ and  $g_2$~modes as a function of the mass of the stellar configuration for hybrid EoSs~1, 2, 3 and 4. For $f$~modes, we use red color for stellar configurations that are stable only in the slow scenario, and black color for stellar configurations that are stable under both scenarios. All the $g$ and $g_2$ modes are presented in red as they are absent if the sharp phase transition is rapid. With dotted line we present the new $g_2$~mode, with continuous line the $g$~mode, and with dash-dotted line the frequency of the $f$~mode. Circles (triangles) over the lines are used to indicate stars with one (two) sharp discontinuities in their cores. For details of the EoS, see table~\ref{tabla:param_selec}.}
  \label{fig:frec-m-dd2}
\end{figure}

\begin{figure}
    \centering
    \includegraphics[width=1.0\columnwidth]{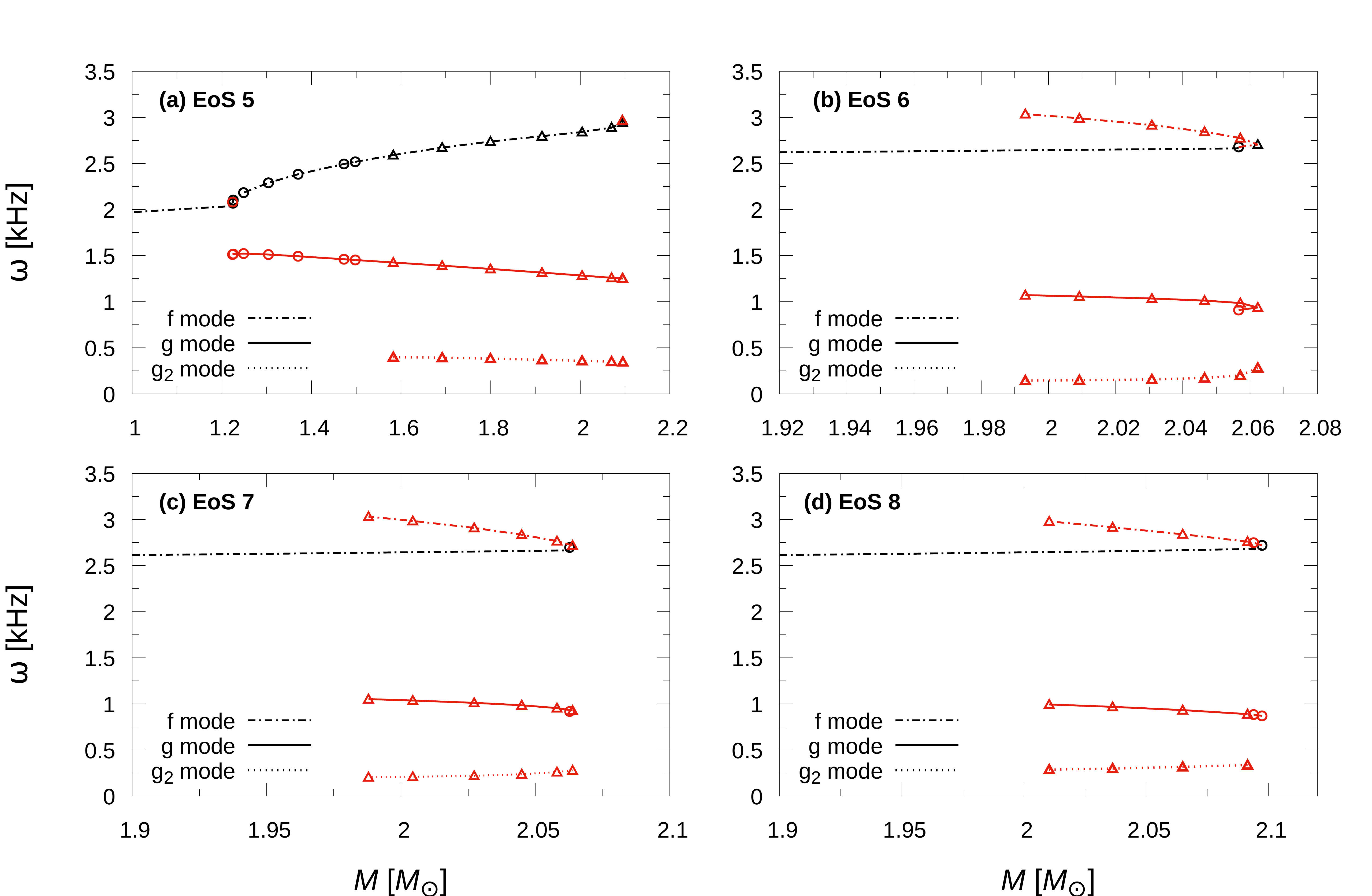}
    \caption{(Color online) Same as Figure \ref{fig:frec-m-dd2} but for hybrid EoSs~5, 6, 7 and 8. For details of the EoSs, see table~\ref{tabla:param_selec}.}
    \label{fig:frec-m-gm1l-sw4l}
\end{figure}

For the hybrid EoSs  in table~\ref{tabla:param_selec}, the frequencies of the normal oscillation modes~$f$ and $g$ were calculated not only for stars with two phase transitions, but also for stars with one phase transition and without phase transition. In figures~\ref{fig:frec-m-dd2} and \ref{fig:frec-m-gm1l-sw4l} we show the frequencies of the oscillation modes, $\omega$, as a function of the stellar configuration mass. To present the results related to the oscillation $f$~modes we use red color for stellar configurations that are stable only in the slow scenario, and black color for stellar configurations that are stable under both scenarios. All the $g$ and $g_2$ modes are presented in red as they are absent if the sharp phase transition is rapid. Each panel corresponds to a different hybrid EoS according to the table~\ref{tabla:param_selec}. The different line types are associated with the different oscillation modes: dash-dotted line for the $f$~mode, continuous line for the $g$~mode and dotted line for the $g_2$~mode, the new mode associated with the abrupt quark-quark phase transition. The different symbols (triangles, hollow dots) refer to stars with different composition: no symbol is used for purely hadronic objects, circles and triangles are used to characterize the models of stars with one and two phase transitions, respectively.

{}

Specifically, in figure~\ref{fig:frec-m-dd2}, we show different frequencies of the oscillation modes as a function of the mass of the compact object. In this figure, we present the results obtained for stellar configurations constructed using the DD2 parametrization. The behavior of EoS~1 and 2 -panel (a) and (b), respectively- is similar and these results are comparable (the only difference between them is a short twin branch of stars with one discontinuity in the EoS~1). In both cases, after the first phase transition there are a disconnected branch that become connected, and so twins configuration appear, when slow conversion is considered. For hybrid EoS~3 -see panel (c) in figure \ref{fig:frec-m-dd2}-, it presents twin configurations and we can observe that the frequency of both $f$ and $g$~modes show a degeneration as a function of the mass. For this reason, a simultaneous detection of both the $f$ and $g$, and (non-)detection of the $g_2$~mode would be needed in order to determine the internal structure of the pulsating compact object. These differences make the frequency of the $f$~mode in pure hadronic stars different from the frequency of the $f$~mode in HSs with one phase transition. Results for hybrid EoS~4 are shown in panel (d). In the rapid scenario, the appearance of quark matter quickly destabilizes stellar configurations and hybrid stellar configurations are only marginal. However, in the slow scenario, an extended stable HS branch of twin configurations appears up to the terminal mass Similar results, for non-parametric hybrid EoSs, were presented in Ref.~\cite{Ranea2018b}.

Analogously, in figure~\ref{fig:frec-m-gm1l-sw4l}, we present the frequencies of the oscillation modes for HSs constructed with GM1L -EoS~5 shown in panel~(a)- and SW4L -EoS~6, 7 and 8 presented in panels~(b), (c) and (d) respectively- parametrizations. Results obtained using EoS~5 are similar to those shown in figure \ref{fig:frec-m-dd2} for hybrid EoSs constructed using DD2. In panel (b), we can see that the branch of stable stars with double phase transition becomes larger when the slow scenario is considered. Panels~(c) and (d) show similar results for hybrid EoSs, in which, in the rapid scenario, the onset of quark matter in the inner core destabilizes the star and a very short branch of connected stable configurations exists. In the slow scenario, stable stars with double phase transition appear in EoS~7 and 8. In the last one, also the branch of stars with one phase transition becomes larger. Regarding these two panels, it is important to remark that the difference between EoS~7 and EoS~8 is the length of the branch of stable stars with one phase transition. For EoS~8, such branch is stable only at the beginning and then it becomes unstable.

In general, stars without phase transition have lower mass. For these models, we have only calculated the fundamental oscillation mode, $f$. For central pressures above $P_1$, stellar configurations with one phase transition appear. In addition to the fundamental mode, for these configurations we have calculated the $g$~mode associated to the discontinuity in the energy density inside the stars. We must emphasize that, if we assume rapid conversion between phases,  EoS~4, 7 and 8 do not produce stable stars with two phase transitions. For stellar configurations with central pressures higher than $P_2$, we obtained stars with double phase transition. For these objects we found, for the first time, a second oscillation mode associated with the second discontinuity in the energy density profile inside the star, we have called this $g_2$~mode. It is important to remark that each mode appears in a particular frequency range. The frequencies calculated for $f$, $g$ and $g_2$~modes are in a range between $2$ to $3$~kHz, $0.8$ to $1.5$~kHz and $0.2$ to $0.5$~kHz, respectively. However, despite the fact that each mode lies in a definite range, for the three modes there are slight but noticeable differences between different EoSs.

\begin{figure}
    \centering
    \includegraphics[width=.8\columnwidth]{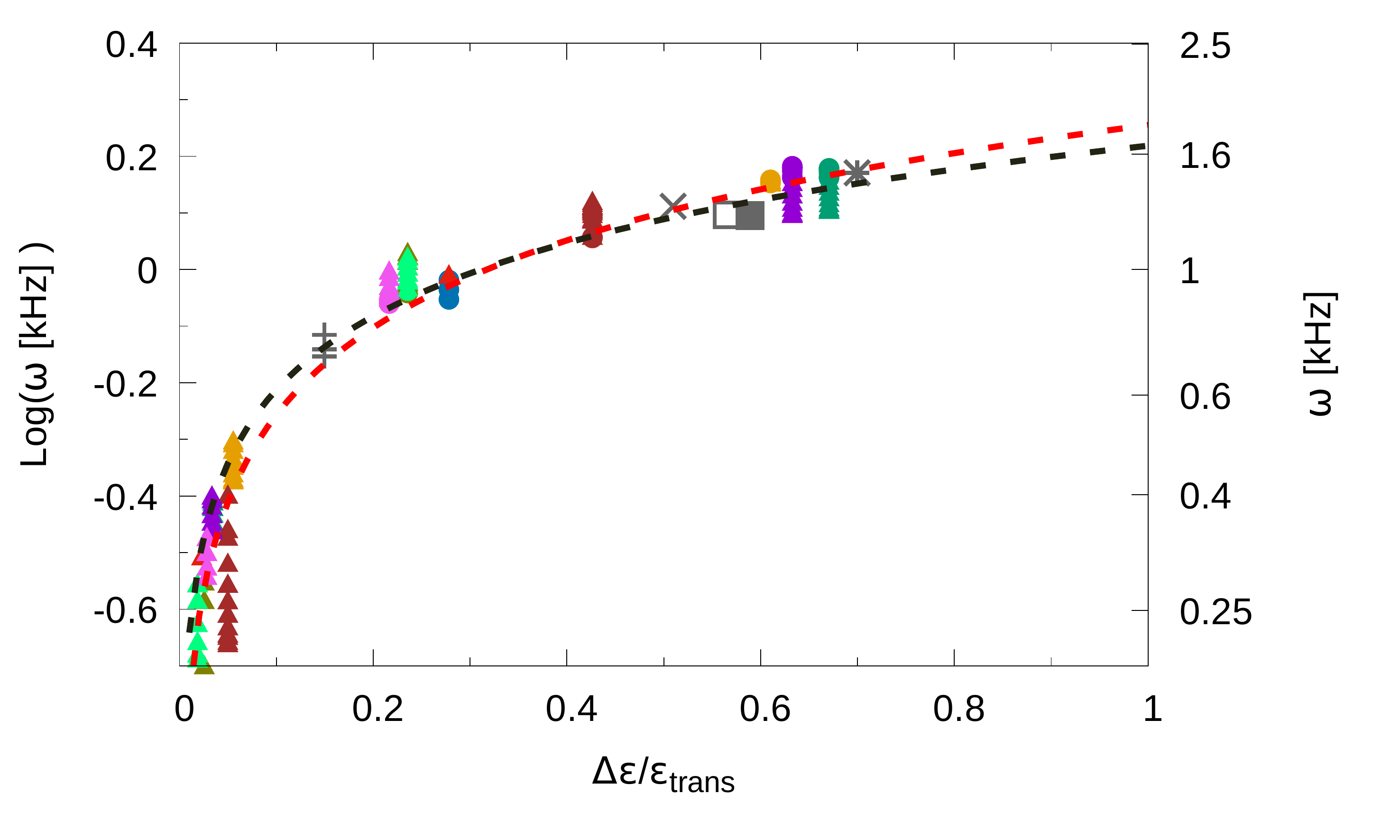}
    \caption{(Color online) In this figure we show the ${\rm Log}(\omega)$ as function of the $\Delta \epsilon_{\rm trans} / \epsilon_{\rm trans}$ for the $g$ and $g_2$~modes. The gray dots and the black curve correspond to the results obtained in Ref.~\cite{Ranea2018b}. In colors, we added the new results obtained in this work, the circles correspond to the $g$~mode of stars with one phase transition and the triangles correspond to the $g$ and the new $g_2$~modes of stars with two phase transition, and the dash-dotted line in red is the new fit which includes all the data. Data grouped in the bottom left corner corresponds to frequencies of $g_2$~modes.}
    \label{fig:delta_e}
\end{figure}

On the other hand, a relation between the frequency of the $g$~mode and the quotient between the jump in the energy density, $\Delta \epsilon$, and the energy density at phase transition, $\epsilon _{\rm trans}$, has been presented in Ref.~\cite{Ranea2018b}. The results obtained in this work for the $g$ and for the $g_2$~mode were added to the frequencies of Ref.~\cite{Ranea2018b}. A new fit for all the data is shown with a red line in figure~\ref{fig:delta_e}. In the same figure, the original fit is shown in black and results from Ref.~\cite{Ranea2018b} are shown with gray color. Modes corresponding to the EoSs used in this work are presented using the same colors of figure \ref{fig:mraio-mrho}. The circles correspond to modes of HSs with one phase transition and the triangles correspond to modes of HSs with two phase transition.

Following Ref.~\cite{Ranea2018b}, we fit the discontinuity modes (both $g$ and $g_2$) using the simple functional form given by,
\begin{equation}
    \omega=a \,{\rm{Log}}\left(\frac{\Delta \epsilon}{\epsilon_{\rm trans}}\right)+b \ ,
\end{equation}{}
where the values of the parameters $a$ and $b$ turn to be  
\begin{align}
    a &= 0.512 \pm 0.022 \quad (4.5\%) \ ,\\
    b &= 0.255 \pm 0.017 \quad (8.1\%).
\end{align}

As it can be seen from figure~\ref{fig:delta_e}, including the new data, represented with colored dots, there is still a strong correlation between the frequency of the discontinuity modes and CSS parameter $\Delta \epsilon / \epsilon _{\rm trans}$. Considering the value of the parameter $a$ resulting after the fit, we have obtained a generalization of the Newtonian case where the value of $a$ is strictly $0.5$, as the square of the Brunt V\"aisl\"al\"a frequency is proportional to $\Delta \epsilon / \epsilon_{\rm trans}$ \cite{McDermott1990}.

Furthermore, we estimate the detectability of the oscillation modes regarding their energetics. In general, for an object at a distance, $D$, the minimum energy that has to be channeled into a particular oscillation mode of frequency, $f$, and quality factor, $Q=\pi f \tau$, with a given signal-to-noise ratio, $S/N$, in order to be detectable by a particular instrument with a noise power spectral density, $S_n$, can be estimated using this simple formula \cite{Kokkotas:1999mn}:

\begin{equation}
\label{energy}
    E_{\rm GW} = 6.25 \times 10^{90}\left(\frac{S}{N}\right)^2\frac{1 +4Q^2}{4Q^2}\left(\frac{D}{10\rm{kpc}}\right)^2\left(\frac{f}{1\rm{kHz}}\right)^2\left(\frac{S_n}{1\rm{Hz}^ {-1}}\right)^2 \rm{erg}.
\end{equation}
For the Advanced LIGO-Virgo detector, $\sqrt{S_n} \sim 10^{-23}$~Hz${}^{-1/2}$ in the frequency band $\sim 0.5$-$2$~kHz. Einstein telescope is expected to be an order of magnitude more sensitive \cite{sensibility}.

In table~\ref{table:energy}, we present minimum values of energy that have to be channeled into the $f$, $g$ and $g_2$~modes in order to be detectable by these two instruments, if the emitting compact object were at a distance of $10$~kpc or $15$~Mpc, and considering a detection threshold of $S/N =8$. To obtain these results, we have used characteristic damping times of $f$-modes and the fact that, for the gravity modes, $Q \gg 1$. Moreover, we have used characteristic values for the $f$, $g$ and $g_2$~modes of $2$, $1$ and $0.5$~kHz, respectively.

\begin{table}[t!]
\centering
\begin{tabular}{lcccc}
\toprule
\multirow{2}{*}{\begin{tabular}[c]{@{}l@{}} \\ Detector \end{tabular}} & \multirow{2}{*}{\begin{tabular}[c]{@{}l@{}} \\ Distance \end{tabular}} & \multicolumn{3}{c}{Energy [erg]} \vspace{0.1cm}  \\
& & $f$-mode & $g$-mode & $g_2$-mode \\
\hline
\midrule
Advanced LIGO-Virgo & 10kpc    &   $2\times 10^{47}$  &    $7\times 10^{46}$     &  $10^{46}$ \\ \midrule
Einstein            & 10kpc    &  $2\times 10^{45}$  &   $7\times 10^{44}$       &      $10^{44}$ \\ \midrule
Advanced LIGO-Virgo & 15Mpc    & $6\times 10^{53}$    &  $9\times 10^{52}$        &    $2\times 10^{52}$ \\ \midrule
Einstein            & 15Mpc    &    $6\times 10^{51}$      &     $9\times 10^{50}$    &    $2\times 10^{50}$ \\
\bottomrule
\end{tabular}
\caption{Estimated values of the minimum amount of energy that needs to be channeled in a particular mode in order to be detected by a particular instrument. Two astrophysical relevant distance scenarios are considered: a Galactic NS (10kpc) and an object at a distance comparable to the Virgo cluster (15Mpc).}
\label{table:energy}
\end{table}

Considering the astrophysical context, a typical core-collapse supernova event releases $\sim 10^{53}$~ erg. In comparison, estimations presented in table~\ref{table:energy} show that detection of GW emitted by Galactic NSs are difficult but feasible even with the Advanced LIGO-Virgo detector, particularly for $g$-modes. Once data from the Einstein telescope become available, this situation would be much more encouraging. The $\sim 10^{45}$~erg detection-threshold for $f$-modes might allow us to observe GW produced as a consequence of the non-radial perturbations of giant flares in magnetars and, even, of glitch activity in local pulsars.


\section{Summary and discussion}
\label{sec:conclusion}

In this work, we have studied the stellar structure and the non-radial oscillations of HSs, considering the occurrence of a second sharp phase transition in their cores. We have considered the scenarios in which conversion between the different phases are rapid and slow, where extended branches of stable stellar configurations exist. We have analyzed the effect of the second transition on the frequencies of the normal oscillation modes $f$ and $g$, within the relativistic Cowling approximation.

To model the two sequential phase transitions we have used the Maxwell construction for the hybrid EoSs. To describe the nuclear matter, we have chosen three modern parametrizations of the RMF theory: GM1L, SW4L and DD2.  Stellar configurations constructed with GM1L and SW4L are compatible with recent astrophysical constraints from event GW170817 and NICER observations of PSR~J0030+0451. On the other hand, stars obtained using DD2 EoSs are not compatible with the restrictions imposed by GW170817. However, our results show that if a hadron-quark phase transition occurs at low pressure, this situation can be overcome. The sequential quark phase transition EoSs have been constructed using an extended generalization of the CSS parametrization. The CSS EoS model has the capability of describing, both qualitative and quantitative, aspects of non-parametric and effective quark matter models of QCD. Moreover, within this approach, observational constraints analysis can be expressed in a model-independent fashion. We have explored, systematically, the six-parameter space of this model. In this way, we have constructed more than 3000 different hybrid EoSs combining different models, exploring differences and similarities among them. The EoSs parameters have been chosen following two cases, stiff and soft, in which the main difference between them is the value of the squared speed of sound. For the stiff case, we use values of the squared speed of sound compatible with the Ref.~\cite{Alford2017}. In the soft case we consider the values of the squared speed of sound compatible with non-parametric quark EoS.

For each hybrid EoS, we have solved TOV equations, from which we obtain stellar parameters such as the mass, $M$, the radius, $R$ and the dimensionless tidal deformability, $\Lambda$. Of all the HSs families obtained, we have selected a set of eight representative hybrid EoSs qualitatively different, guaranteeing that the $ \sim 2M_\odot$ constraint is satisfied. An interesting fact is that, in the rapid conversion scenario, despite having analyzed $\sim 3400$ different hybrid EoSs and having obtained long branches of both connected and disconnected hybrid stellar configurations, our models only predict the appearance of twin stars in a very restrictive range of masses. This situation might occur for low mass objects as in EoSs~1 and 5 or, as in EoS~3, for objects with masses $\sim 2M_\odot$. Another interesting remark is that we have not found triplet configurations like those found in Refs.~\cite{Alford2017,Li:2019}. Due to the lack of hybrid EoSs, that lead to the appearance of triplets, and the extremely short mass-range of the triplet family found in Refs.~\cite{Alford2017,Li:2019}, we argue that this is a relatively marginal feature. In any case, their appearance is a product of a fine-tuned selection of EoS parameters. This situation changes when we consider slow phase transitions. In this theoretical scenario, high-mass slow twins are a common feature. Moreover, we have found that slow triplets are also possible.
 
The conclusions arising from the results of the dimensionless tidal deformability, $\Lambda$, are similar to those obtained from mass radius relation. The GM1L and SW4L parametrizations satisfy the constraints coming from the GW170817 event, but only the low pressure transition case for HSs constructed with DD2 parametrization does it. Regarding this observable, the major differences between the slow and rapid conversion scenarios occur for HSs with masses above $\sim 2M_\odot$. For this reason, the data from GW170817 is not useful to decide which of these two theoretical possibilities could be favored if a phase transition occurs in the inner cores of HSs.

In summary, we have constructed HSs with sequential phase transitions satisfying recent restrictions from neutron stars observations, even considering hadronic EoSs that, otherwise, should be discarded. It is important to remark that, we have selected a single soft case as representative of all the other soft cases, since they are very similar and do not have differentiating characteristics. The situation is different for the stiff cases. However, most of the quark EoSs used in the literature to describe the matter at high densities in HSs (non-parametric quark EoS) have a squared speed of sound of the order of $\sim~0.3$, \emph{i.e} they are soft EoSs. This causes that the appearance of twins and triplets is even more unlikely, since the favored scenario for the formation of such configurations is a stiff high density EoS (like the parametric EoSs used in this work). It is known that by including repulsive interactions in the quark models, the EoS becomes stiffen. In this sense, the parametric quark EoSs with large speed of sound could be associated to more realistic quark EoSs in which higher order repulsive interactions are included \cite{Providencia2020}.

In addition, for the selected EoSs, we have calculated the frequencies for the non-radial $f$ and $g$~modes, the latter associated with sharp and slow phase transitions inside HSs. To this aim, we have extended the Cowling approximation to consider sequential phase transitions. We found $f$~modes to lie in the range $2$-$3$~kHz and $g$~modes in the $0.8$-$1.5$~kHz range. However, it is important to point out that the differences for the $f$~modes when using the Cowling approximation could be up to $20\%$ when compared to the General Relativity linearized equations. For this reason, our conclusions regarding the $f$~modes are only qualitative. Nevertheless, as for the $g$~modes the differences are less than the $10\%$, we could assert that the detection of GW associated with an oscillation mode with a frequency in the range of $1$-$1.5$~kHz  would have important astrophysics implications: it might be not only a proof of the existence of quark matter inside a compact star, but also it might indicate that the hadron-quark transition phase is abrupt, imposing constraints to the surface tension value at the hadron-quark interface.

Additionally, the occurrence of a second abrupt phase transition in HSs, leads to the appearance of a new $g$~mode, called $g_2$~mode, which has not been calculated before. This mode typically appears with lower frequencies than the $g$~modes, in the range $0.2$-$0.5$~kHz. Furthermore, the relation shown in figure~\ref{fig:delta_e} strengthen the idea presented in Ref.~\cite{Ranea2018b} that the frequencies of the $g$-modes, both $g$ and the new $g_2$, are universally related to the value of $\Delta \epsilon /\epsilon _{\rm trans}$. For this reason a detection of such modes might be the key to properly understand the nature of the hadron-quark phase transition and restrict some of its physical parameters associated with NSs internal composition.

The $g_2$-modes presented in this work lies in a frequency range that might be resonantly excited by tidal forces during the early stages of a NS-NS (BH-NS) inspiral, when orbital period might (momentarily) become resonant with such modes (see, for example, Refs. \cite{resonant01,resonant02}). During this time, energy is drawn from the orbital motion to excite such modes and as a consequence the inspiral speeds up. This effect might produce an observable phase shift in the gravitational wave forms that could, in principle, allow us to shed some light into the nature of the innermost regions of these extremely compact objects.

It is important to point out that in the $g_2$~mode range of frequencies, there are different modes associated to other effects such as superfluidity, rotation and intense magnetic field inside NSs \cite{Miniutti2003, Lugones2016}. Therefore, as GW information would not be sufficient to distinguish between certain modes, we need, also, information from electromagnetic observations to discard modes in that frequency range produced by physical effects other than a second sharp phase transition. In this way, we could properly identify the different oscillation modes whose frequencies are similar and characterize the emitting object. This reinforces the fact that multimessenger astronomy is the most suitable approach to better understand the internal composition of such compact objects.

It would be extremely interesting to analyze the existence of a universal relationship like the one presented in figure~\ref{fig:delta_e} beyond the relativistic Cowling approximation. Studying $g$~mode oscillations within the linearized theory of relativity would allow to investigate not only the oscillation frequency but also the damping time. If this were  possible, it would have a great impact as a new branch of study would open: the $g$~mode asteroseismology.

To conclude, all this rich complexity in the QNM spectrum might be tested, as we have estimated in table~\ref{table:energy}, when observational data from third generation GW-observatories become available. During this era, the study of such modes might be useful to shed some light into understanding the nature of matter in the inner cores of such compact objects.


\acknowledgments
MCR is a fellow of CONICET. MCR, IFR-S, MM, GM and MO thank CONICET and UNLP for financial support under grants PIP-0714, G140 and G157. OMG is partially support by UNLP grant G157. OMG acknowledges the hosting by IA-PUC as an invited researcher. The authors thank the anonymous referee for the constructive comments that have contributed to improve the quality of the manuscript.




\begin{thebibliography}{99}

\bibitem{Ozel2016}
F.~Ozel and P.~Freire, \emph{Masses, Radii, and the Equation of State of Neutron Stars}, \emph{Annual Review of Astronomy and Astrophysics} {\bf 54} (2016), 401-440.

\bibitem{Steiner2018}
A.~W.~Steiner, C.~O.~Heinke, S.~Bogdanov, C.~K.~Li, W.~C.~G.~Ho, A.~Bahramian and S.~Han, \emph{Constraining the mass and radius of neutron stars in globular clusters}, \emph{Monthly Notices of the Royal Astronomical Society} {\bf 476} (2018), 421-435.

\bibitem{Fridolin1999}
F.~Weber, \emph{Pulsars as astrophysical laboratories for nuclear and particle physics}, \emph{IOP Publishing, Bristol, U.K.} (1999).

\bibitem{Orsaria2019}
M.~Orsaria, G.~Malfatti, M.~Mariani, I.~Ranea-Sandoval, F.~Garc{\'{\i}}a, W.~Spinella, G.~Contrera, G.~Lugones and F.~Weber, \emph{Phase transitions in neutron stars and their links to gravitational waves}, \emph{Journal of Physics G: Nuclear and Particle Physics} {\bf 46} (2019), 073002.

\bibitem{Ranea2018b}
I.~F.~Ranea-Sandoval, O.~M.~Guilera, M.~Mariani and M.~G.~Orsaria, \emph{Oscillation modes of hybrid stars within the relativistic Cowling approximation}, \emph{Journal of Cosmology and Astroparticle Physics} {\bf 2018} (2018), 031.

\bibitem{Alford2013}
M.~Alford, S.~Han and M.~Prakash, \emph{Generic conditions for stable hybrid stars}, \emph{Physical Review D: Particles, Fields, Gravitation and Cosmology} {\bf 88} (2013), 083013.

\bibitem{Lugones2016}
G.~Lugones, \emph{From quark drops to quark stars. Some aspects of the role of quark matter in compact stars}, \emph{European Physical Journal A} {\bf 52} (2016), 53.

\bibitem{Blaschke2014}
D.~B.~Blaschke, H.~A.~Grigorian, D.~E.~Alvarez-Castillo and A.~S.~Ayriyan, \emph{Mass and radius constraints for compact stars and the QCD phase diagram}, \emph{Journal of Physics Conference Series} {\bf 496} (2014), 012002.

\bibitem{Benic2015}
S.~Beni{\'c}, D.~Blaschke, D.~E.~Alvarez-Castillo, T.~Fischer and S.~Typel, \emph{A new quark-hadron hybrid equation of state for astrophysics. I. High-mass twin compact stars}, \emph{Astronomy \& Astrophysics} {\bf 577} (2015), A40.
\bibitem{Alford2017}
M.~Alford and A.~Sedrakian, \emph{Compact Stars with Sequential QCD Phase Transitions}, \emph{Physical Review Letters} {\bf 119} (2017), 161104.

\bibitem{Li:2019}
J.~J.~Li, A.~Sedrakian and M.~Alford, \emph{Relativistic hybrid stars with sequential first-order phase transitions and heavy-baryon envelopes}, \emph{Phys.~Rev.~D} {\bf 101} (2020), no.~6, 063022.

\bibitem{Alford:2008}
M.~Alford, A.~Schmitt, K.~Rajagopal and T.~Sch{\"a}fer, \emph{Color superconductivity in dense quark matter}, \emph{Reviews of Modern Physics} {\bf 80} (2008), 1455-1515.

\bibitem{Ranea2016}
I.~Ranea-Sandoval, S.~Han, M.~Orsaria, G.~Contrera, F.~Weber and M.~Alford, \emph{Constant-sound-speed parametrization for Nambu-Jona-Lasinio models of quark matter in hybrid stars}, \emph{Physical Review C: Nuclear Physics} {\bf 93} (2016), 045812.

\bibitem{Demorest2010}
P.~Demorest, T.~Pennucci, S.~Ransom, M.~Roberts and J.~Hessels, \emph{A two-solar-mass neutron star measured using Shapiro delay}, \emph{Nature} {\bf 467} (2010), 1081-1083.

\bibitem{antoniadis2013}
J.~Antoniadis et al., \emph{A Massive Pulsar in a Compact Relativistic Binary}, \emph{Science} {\bf 340} (2013), 448.

\bibitem{Cromartie2019}
H.~Cromartie et al., \emph{Relativistic Shapiro delay measurements of an extremely massive millisecond pulsar}, \emph{Nature Astronomy} {\bf 4} (2020), 72-76.

\bibitem{Lattimer2004}
J.~Lattimer and M.~Prakash, \emph{The Physics of Neutron Stars}, \emph{Science} {\bf 304} (2004), 536-542.

\bibitem{Lattimer2014}
J.~Lattimer and A.~W.~Steiner, \emph{Neutron Star Masses and Radii from Quiescent Low-mass X-Ray Binaries}, \emph{The Astrophysical Journal} {\bf 784} (2014), 123.

\bibitem{Andersson2011}
N.~Andersson et al., \emph{Gravitational waves from neutron stars: promises and challenges}, \emph{General Relativity and Gravitation} {\bf 43} (2011), 409-436.

\bibitem{Lasky:2015}
P.~D.~Lasky, \emph{Gravitational Waves from Neutron Stars: A Review}, \emph{Publ.~Astron.~Soc.~of Australia} {\bf 32} (2015), e034.

\bibitem{Abbott2017b}
B.~P.~Abbott et al., \emph{Multi-messenger Observations of a Binary Neutron Star Merger}, \emph{The Astrophysical Journal, Letters} {\bf 848} (2017), L12.

\bibitem{Abbott2017a}
B.~P.~Abbott et al., \emph{GW170817: Observation of Gravitational Waves from a Binary Neutron Star Inspiral}, \emph{Phys.~Rev.~Lett.} {\bf 119} (2017), 161101.

\bibitem{Raithel2018}
C.~Raithel, F.~Özel and D.~Psaltis, \emph{Tidal deformability from GW170817 as a direct probe of the neutron star radius}, \emph{Astrophys.~J.~Lett.} {\bf 857} (2018), no.~2, L23.

\bibitem{Annala2018}
E.~Annala, T.~Gorda, A.~Kurkela and A.~Vuorinen, \emph{Gravitational-Wave Constraints on the Neutron-Star-Matter Equation of State}, \emph{Physical Review Letters} {\bf 120} (2018), 172703.

\bibitem{quark-NS}
E.~Annala, T.~Gorda, A.~Kurkela, J.~Nättilä and A.~Vuorinen, \emph{Evidence for quark-matter cores in massive neutron stars}, \emph{Nature Phys.} (2020).

\bibitem{Ligo2020}
B.~P.~Abbott et al., \emph{GW190425: Observation of a Compact Binary Coalescence with Total Mass $\sim 3.4 M_{\odot}$}, \emph{The Astrophysical Journal} {\bf 892} (2020), L3.

\bibitem{Riley2019}
T.~E.~Riley et al., \emph{A NICER View of PSR~J0030+0451: Millisecond Pulsar Parameter Estimation}, \emph{The Astrophysical Journal, Letters} {\bf 887} (2019), L21.

\bibitem{Miller2019}
M.~C.~Miller et al., \emph{PSR~J0030+0451 Mass and Radius from NICER Data and Implications for the Properties of Neutron Star Matter}, \emph{The Astrophysical Journal, Letters} {\bf 887} (2019), L24.

\bibitem{Glampedakis:2018}
K.~Glampedakis and L.~Gualtieri, \emph{Gravitational waves from single neutron stars: an advanced detector era survey}, \emph{Astrophys.~Space Sci.~Libr.} {\bf 457} (2018), 673-736.

\bibitem{Andersson1998}
N.~Andersson and D.~Kokkotas K.~, \emph{Towards gravitational wave asteroseismology}, \emph{Monthly Notices of the Royal Astronomical Society} {\bf 299} (1998), 1059-1068.

\bibitem{extended-stability} J.~P. Pereira, C.~V. Flores, G. Lugones, \emph{Phase Transition Effects on the Dynamical Stability of Hybrid Neutron Stars}, ApJ {\bf{860}}, 1, 12 (2018).

\bibitem{Stephanov:2005} M.~A.~Stephanov, \emph{QCD phase diagram and the critical point}, Prog.Theor.Phys.Suppl.{\bf 153} (2004), 139-156; Int.J.Mod.Phys.A {\bf 20} (2005), 4387-4392.

\bibitem{Caines:2017} H.~Caines, \emph{The Search for Critical Behavior and Other Features of the QCD Phase Diagram – Current Status and Future Prospects}, Nuclear Physics A {\bf 967} (2017),  121.

\bibitem{Chodos1974}
A.~Chodos, R.~L.~Jaffe, K.~Johnson, C.~B.~Thorn and V.~F.~Weisskopf, \emph{New extended model of hadrons}, \emph{Physical Review D: Particles, Fields, Gravitation and Cosmology} {\bf 9} (1974), 3471-3495.

\bibitem{Contrera2010}
G.~A.~Contrera, M.~Orsaria and N.~N.~Scoccola, \emph{Nonlocal Polyakov-Nambu-Jona-Lasinio model with wave function renormalization at finite temperature and chemical potential}, \emph{Physical Review D: Particles, Fields, Gravitation and Cosmology} {\bf 82} (2010), 054026.

\bibitem{Orsaria2013}
M.~Orsaria, H.~Rodrigues, F.~Weber and G.~A.~Contrera, \emph{Quark-hybrid matter in the cores of massive neutron stars}, \emph{Physical Review D: Particles, Fields, Gravitation and Cosmology} {\bf 87} (2013), 023001.

\bibitem{Ranea2019}
I.~F.~Ranea-Sandoval, M.~Orsaria, G.~Malfatti, D.~Curin, M.~Mariani, G.~A.~Contrera and O.~M.~Guilera, \emph{Effects of Hadron-Quark Phase Transitions in Hybrid Stars within the NJL Model}, \emph{Symmetry} {\bf 18} (2019), 425.

\bibitem{Nefediev2009}
A.~V.~Nefediev, Y.~A.~Simonov and M.~A.~Trusov, \emph{Deconfinement and Quark-Gluon Plasma}, \emph{International Journal of Modern Physics E} {\bf 11} (2019), 549-599.

\bibitem{Mariani2017}
M.~Mariani, M.~Orsaria and H.~Vucetich, \emph{Constant entropy hybrid stars: a first approximation of cooling evolution}, \emph{Astronomy \& Astrophysics} {\bf 601} (2017), A21.

\bibitem{Mariani2019}
M.~Mariani, M.~G.~Orsaria, I.~F.~Ranea-Sandoval and O.~M.~Guilera, \emph{Hybrid magnetized stars within the Field Correlator Method}, \emph{Boletin de la Asociacion Argentina de Astronomia La Plata Argentina} {\bf 61} (2019), 231-233.

\bibitem{Voskresensky2003}
D.~N.~Voskresensky, M.~Yasuhira and T.~Tatsumi, \emph{Charge screening at first order phase transitions and hadron-quark mixed phase}, \emph{Nuclear Physics A} {\bf 723} (2003), 291-339.

\bibitem{Pinto2012}
M.~B.~Pinto, V.~Koch and J.~Randrup, \emph{Surface tension of quark matter in a geometrical approach}, \emph{Physical Review C: Nuclear Physics} {\bf 86} (2012), 025203.

\bibitem{Lugones2013}
G.~Lugones, A.~G.~Grunfeld and M.~Al Ajmi, \emph{Surface tension and curvature energy of quark matter in the Nambu-Jona-Lasinio model}, \emph{Physical Review C: Nuclear Physics} {\bf 88} (2013), 045803.

\bibitem{BPS} G. Baym,J.~C. Pethick and P. Sutherland, \emph{The Ground State of Matter at High Densities: Equation of State and Stellar Models}, The Astrophysical Journal {\bf 170}, (1971), 299.

\bibitem{BBP} G. Baym, H.~A. Bethe and J.~C. Pethick, \emph{Neutron star matter}, \ Nuclear Physics A {\bf 175}, (1971) 225–271.

\bibitem{crust-fortin} M. Fortin et. al, \emph{Neutron star radii and crusts: Uncertainties and unified equations of state}, Physical Review C {\bf{94}}, 035804 (2016)
%

\bibitem{Spinella2017}
W.~M.~Spinella, \emph{A Systematic Investigation of Exotic Matter in Neutron Stars}, \emph{The Claremont Graduate University} (2017).

\bibitem{Typel2010}
S.~Typel, G.~R{\"o}pke, T.~Kl{\"a}hn, D.~Blaschke and H.~H.~Wolter, \emph{Composition and thermodynamics of nuclear matter with light clusters}, \emph{Physical Review C} {\bf 81} (2010), 015803.

\bibitem{Spinella:2018dab}
W.~M.~Spinella and F.~Weber, \emph{Hyperonic Neutron Star Matter in Light of GW170817}, \emph{Astron.~Nachr.} {\bf 340} no.~1-3, (2019), 145-150.

\bibitem{Ranea2017}
I.~F.~Ranea-Sandoval, M.~G.~Orsaria, S.~Han, F.~Weber and W.~M.~Spinella, \emph{Color superconductivity in compact stellar hybrid configurations}, \emph{Physical Review C: Nuclear Physics} {\bf 96} (2017), 065807.

\bibitem{Malfatti2019}
G.~Malfatti, M.~G.~Orsaria, G.~A.~Contrera, F.~Weber and I.~F.~Ranea-Sandoval, \emph{Hot quark matter and (proto-) neutron stars}, \emph{Physical Review C: Nuclear Physics} {\bf 100} (2019), 015803.

\bibitem{Malfatti2020}
G.~Malfatti, M.~G.~Orsaria, I.~F.~Ranea-Sandoval, G.~A.~Contrera and F.~Weber, \emph{Delta baryons and diquark formation in the cores of neutron stars}, Phys. Rev D {\bf{102}}, 063008, (2020).

\bibitem{BCS}
{J.~Bardeen, L.~N.~Cooper, and J.~R.~Schrieffer, \emph{Theory of Superconductivity}, \emph{Phys.~Rev}. {\bf{106}} (1957), 162; \emph{Microscopic Theory of Superconductivity}, \emph{Phys.~Rev.} {\bf{108}} (1957), 1175}.

\bibitem{Shovkovy_2005}
{I.~Shovkovy, \emph{Two Lectures on Color Superconductivity},\emph{ Found. Phys.} {\bf{35}}, (2005), 1309.}

\bibitem{Buballa_2005}
{M. Buballa, \emph{NJL-model analysis of dense quark matter}, \emph{Phys. Rept.} {\bf{407}} (2005), 205.}

\bibitem{Castillo2017}
D.~E.~ Alvarez-Castillo and D.~B.~Blaschke,  \emph{High-mass twin stars with a multipolytrope equation of state}, \emph{Phys.~Rev.~C} {\bf 96} (2017), 045809.

\bibitem{Montana2019}
G.~Montana, L.~Tolos, M.~Hanauske, L.~Rezzolla, \emph{Constraining twin stars with GW170817}, \emph{Phys.~Rev.~D} {\bf 99} (2019), 103009.

\bibitem{Bhattacharyya:2020paf}
S.~Bhattacharyya, \emph{The permanent ellipticity of the neutron star in PSR~J1023+0038}, \emph{Monthly Notices of the Royal Astronomical Society} {\bf 8} (2020).

\bibitem{Abbott:2020lqk}
R.~Abbott et al. [LIGO Scientific and Virgo], \emph{Gravitational-wave constraints on the equatorial ellipticity of millisecond pulsars}, \emph{arXiv e-prints} (2020), arXiv:2007.14251.

\bibitem{tolman}
R.~C.~Tolman, \emph{Static Solutions of Einstein's Field Equations for Spheres of Fluid}, \emph{Phys.~Rev.} {\bf 55} (1939), 364.

\bibitem{OV}
J.~R.~Oppenheimer and G.~M.~Volkoff, \emph{On Massive Neutron Cores}, \emph{Phys.~Rev.} {\bf 55} (1939), 374.

\bibitem{2020PhRvD.102b8501T} Tak{\'a}tsy, J., Kov{\'a}cs, P.\ \emph{Comment on ``Tidal Love numbers of neutron and self-bound quark stars''}, Physical Review D {\bf{102}}, 028501, (2020)

\bibitem{Gerlach:1968}
U.~H.~Gerlach, \emph{Equation of State at Supranuclear Densities and the Existence of a Third Family of Superdense Stars}, \emph{Phys.~Rev.} {\bf 172} (1968), 1325-1330.
\bibitem{Schertler:2000}
K.~Schertler et al., \emph{Quark phases in neutron stars and a 'third family' of compact stars as a signature for phase transitions}, \emph{Nucl.~Phys.~A} {\bf 677} (2000), 463-490.

\bibitem{Banik:2002}
S.~Banik and D.~Bandyopadhyay, \emph{Color superconducting quark matter core in the third family of compact stars}, \emph{Phys.~Rev.~D} {\bf 67} (2003), 123003.


\bibitem{Sieniawska:2018}
M.~Sieniawska et al., \emph{Tidal deformability and other global parameters of compact stars with strong phase transitions}, \emph{Astron.~Astrophys.} {\bf 622} (2019), A174.

\bibitem{Alvarez-Castillo:2018}
D.~E.~Alvarez-Castillo et al., \emph{Third family of compact stars within a nonlocal chiral quark model equation of state}, \emph{Phys.~Rev.~D} {\bf 99} (2019), no.~6, 063010.

\bibitem{pasta-twins1}
A.~Ayriyan et al., \emph{Robustness of third family solutions for hybrid stars against mixed phase effects}, \emph{Phys.~Rev.~C} {\bf 97} (2018), no.~4, 045802.

\bibitem{pasta-twins2}
K.~Maslov et al., \emph{Hybrid equation of state with pasta phases and third family of compact stars}, \emph{Phys.~Rev.~C} {\bf 100} (2019), no.~2, 025802.

\bibitem{kk-and-2002}
K.~D.~Kokkotas and N.~Andersson, \emph{Oscillation and instabilities of relativistic stars}, \emph{Recent Developments in General Relativity} (2002), 121–139.

\bibitem{Kokkotas1999}
K.~D.~Kokkotas and B.~G.~Schmidt, \emph{Quasi-Normal Modes of Stars and Black Holes}, \emph{Living Reviews in Relativity} {\bf 2} (1999), 2.

\bibitem{Kokkotas1992}
K.~D.~Kokkotas and B.~F.~Schutz, \emph{W-modes - A new family of normal modes of pulsating relativistic stars}, \emph{Monthly Notices of the Royal Astronomical Society} {\bf 255} (1992), 119-128.

\bibitem{Finn1987}
L.~S.~Finn, \emph{G-modes in zero-temperature neutron stars}, \emph{Monthly Notices of the Royal Astronomical Society} {\bf 227} (1987), 265-293.


\bibitem{tonetto2020}
L.~Tonetto and G.~Lugones, \emph{Discontinuity gravity modes in hybrid stars: assessing the role of rapid and slow phase conversions}, \emph{Phys.~Rev.~D} \textbf{101} (2020), no.~12, 123029.

\bibitem{Miniutti2003}
G.~Miniutti ,J.~A.~Pons, E.~Berti, L.~Gualtieri and V.~Ferrari, \emph{Non-radial oscillation modes as a probe of density discontinuities in neutron stars}, \emph{Monthly Notices of the Royal Astronomical Society} {\bf 338} (2003), 389-400.

\bibitem{Finn1988}
L.~S.~Finn, \emph{Relativistic stellar pulsations in the Cowling approximation}, \emph{Monthly Notices of the Royal Astronomical Society} {\bf 232} (1988), 259-275.

\bibitem{VasquezFlores2014}
C.~V{\'a}squez Flores and G.~Lugones, \emph{Discriminating hadronic and quark stars through gravitational waves of fluid pulsation modes}, \emph{Classical and Quantum Gravity} {\bf 31} (2014), 155002.

\bibitem{Chirenti2015}
C.~Chirenti, G.~H.~de Souza and W.~Kastaun, \emph{Fundamental oscillation modes of neutron stars: Validity of universal relations}, \emph{Physical Review D: Particles, Fields, Gravitation and Cosmology} {\bf 31} (2014), 155002.

\bibitem{Sotani2002}
H.~Sotani, N.~Yasutake, T.~Maruyama and T.~Tatsumi, \emph{Signatures of hadron-quark mixed phase in gravitational waves}, \emph{Physical Review D: Particles, Fields, Gravitation and Cosmology} {\bf 65} (2002), 024010.

\bibitem{Detweiler1985}
S.~Detweiler and L.~Lindblom, \emph{On the nonradial pulsations of general relativistic stellar models}, \emph{The Astrophysical Journal} {\bf 292} (1985), 12-15.

\bibitem{Sotani2011}
H.~Sotani, K.~Tominaga and K.~Maeda, \emph{Density discontinuity of a neutron star and gravitational waves}, \emph{Physical Review D: Particles, Fields, Gravitation and Cosmology} {\bf 83} (2011), 024014.

\bibitem{McDermott1990}
P.~N.~McDermott, \emph{Density Discontinuity G-Modes}, \emph{Monthly Notices of the Royal Astronomical Society} {\bf 245} (1990), 508.

\bibitem{Kokkotas:1999mn}
K.~D.~Kokkotas, T.~A.~Apostolatos and N.~Andersson, \emph{The Inverse problem for pulsating neutron stars: A 'Fingerprint analysis' for the supranuclear equation of state}, \emph{Mon.~Not.~Roy.~Astron.~Soc.} {\bf 320} (2001), 307-315.

\bibitem{sensibility}
B.~P.~Abbott et al. [LIGO Scientific], \emph{Exploring the Sensitivity of Next Generation Gravitational Wave Detectors}, \emph{Class.~Quant.~Grav.} {\bf 34} (2017), no.~4, 044001.

\bibitem{Providencia2020}
M.~Ferreira, R.~C.~Pereira and C.~Providencia, \emph{Quark matter in light neutron stars}, \emph{arXiv e-prints} (2020), arXiv:2008.12563.

\bibitem{resonant01} A.~Reisenegger,P.~ Goldreich, \emph{Excitation of Neutron Star Normal Modes during Binary Inspiral}, The Astrophysical Journal {\bf 426} (1994), 688.

\bibitem{resonant02} W.~Xu, D.~Lai, \emph{Resonant tidal excitation of oscillation modes in merging binary neutron stars: Inertial-gravity modes}, Phys. Rev. D {\bf 96}, (2017), 083005.
%
\end{thebibliography}
\end{document}